\documentclass[aps,prb,reprint,superscriptaddress,longbibliography]{revtex4-2}

\usepackage{amsmath}
\usepackage{amssymb}
\usepackage{bm}
\usepackage{graphicx}
\usepackage{siunitx}
\usepackage{algpseudocode}
\usepackage{microtype}  
\usepackage[hidelinks]{hyperref}

\newcounter{algorithm}
\renewcommand{\thealgorithm}{S\arabic{algorithm}}
\newenvironment{algorithm}[1][]{%
  \par\addvspace{\baselineskip}%
  \refstepcounter{algorithm}%
  \hrule\vskip3pt
  \def\caption##1{\noindent\textbf{Algorithm~\thealgorithm.} ##1\par\vskip3pt\hrule\vskip3pt}%
}{%
  \vskip3pt\hrule\par\addvspace{\baselineskip}%
}

\newcommand{\startsupplement}{%
  \setcounter{section}{0}%
  \setcounter{equation}{0}%
  \setcounter{figure}{0}%
  \setcounter{table}{0}%
  \renewcommand{\thesection}{S\arabic{section}}%
  \renewcommand{\thesubsection}{\thesection.\arabic{subsection}}%
  \renewcommand{\theequation}{S\arabic{equation}}%
  \renewcommand{\thefigure}{S\arabic{figure}}%
  \renewcommand{\thetable}{S\arabic{table}}%
  \renewcommand{\theHsection}{S\arabic{section}}%
  \renewcommand{\theHsubsection}{S\arabic{section}.\arabic{subsection}}%
  \renewcommand{\theHequation}{S\arabic{equation}}%
  \renewcommand{\theHfigure}{S\arabic{figure}}%
  \renewcommand{\theHtable}{S\arabic{table}}%
}

\begin{document}

\title{The Deformed Image Vortex Ansatz: A Perturbation-Aware Description of Magnetic Vortices in In-Plane Fields}

\author{Thomas G. Coppée}
\email{thomas.coppee@uclouvain.be}
\affiliation{Institute of Condensed Matter and Nanosciences, Université catholique de Louvain, Place Croix du Sud 1, 1348 Louvain-la-Neuve, Belgium}

\author{Colin Ducarme}
\affiliation{Institute of Condensed Matter and Nanosciences, Université catholique de Louvain, Place Croix du Sud 1, 1348 Louvain-la-Neuve, Belgium}

\author{Simon de Wergifosse}
\affiliation{Univ. Grenoble Alpes, CEA, CNRS, Grenoble INP, SPINTEC, 38000 Grenoble, France}

\author{Flavio Abreu Araujo}
\email[Corresponding author: ]{flavio.abreuaraujo@uclouvain.be}
\affiliation{Institute of Condensed Matter and Nanosciences, Université catholique de Louvain, Place Croix du Sud 1, 1348 Louvain-la-Neuve, Belgium}

\begin{abstract}

Thiele-based descriptions of magnetic vortex dynamics in thin ferromagnetic nanodots rely on magnetization ansätze that describe the equilibrium texture but cannot represent perturbation-induced deformations.
We introduce the Deformed Image Vortex Ansatz (DIVA): a perturbation-aware ansatz in which the response to an external perturbation is built into the magnetization profile itself, rather than appended to the dynamics as a correction. Here, we demonstrate the concept for a uniform, stationary in-plane field applied to a Permalloy nanodot, for which the deformation is analytically tractable.
A symmetry-based perturbative expansion identifies the leading deformation as a single $m = 1$ harmonic around the disk, while energy minimization and a dominant-balance analysis yield a closed-form interpolant for the radial profile.
Benchmarked against micromagnetic simulations on Permalloy disks of aspect ratio $t/R = 0.1$ and $0.0125$, this realization reduces the disk-averaged angular deviation by a factor of 3 to 6 relative to the two-vortex ansatz, depending on geometry and field, and reduces the total-energy deviation by about a factor of six in the thicker disk.
\end{abstract}

\maketitle

\section{Introduction}

Magnetic vortices are topologically nontrivial spin textures stabilized in soft ferromagnetic nanodots by the competition between exchange and magnetostatic interactions: the in-plane magnetization curls around a nanometric region in which it tilts out of plane to avoid an exchange singularity \cite{shinjo2000, wachowiak2002}. Two topological characteristics fully represent this configuration: the chirality $C = \pm 1$, corresponding to the sense of rotation of the in-plane magnetization, and the polarity $P = \pm 1$, corresponding to the orientation of the core magnetization.

Their low-frequency gyrotropic motion, in which the core orbits the disk center under the balance of gyrotropic, restoring, dissipative and spin-torque forces, forms the basis of spin-transfer vortex oscillators (STVOs), where a spin-polarized current sustains the orbit and converts direct current injection into a microwave signal \cite{pribiag2007, dussaux2010}. STVOs are intrinsically nonlinear devices because both the orbit radius and oscillation frequency depend on the instantaneous vortex core position. This combination of low phase noise, substantial output power, wide tunability, and rich nonlinearity has positioned STVOs as nanoscale microwave sources and as promising building blocks for neuromorphic computing architectures \cite{torrejon2017, romera2018}.

The dynamics of the vortex core is commonly described within the Thiele formalism, a reduced-order approach in which the full micromagnetic problem is collapsed onto the trajectory $\mathbf{X}(t)$ of the core position via a balance of forces:
\begin{equation}
    \mathbf{G} \times \dot{\mathbf{X}} - \mathbf{D}\dot{\mathbf{X}} - \nabla_{\mathbf{X}}U = \mathbf{F}_{\text{ext}},
\label{eq:thiele}
\end{equation}
where $\mathbf{G}$ is the gyrovector arising from the vortex topology, $\mathbf{D}$ the damping tensor, $U$ the potential energy, and $\mathbf{F}_{\text{ext}}$ the sum of all external forces such as Zeeman or spin-transfer force \cite{thiele1973, guslienko2006}. All dynamical coefficients are obtained by projecting the Landau--Lifshitz--Gilbert--Slonczewski equation onto a prescribed magnetization profile, so that the predictive accuracy of every Thiele-based model is set by the quality of this underlying ansatz \cite{dewergifosse2023}.

Existing analytical descriptions therefore rely on a hierarchy of increasingly detailed magnetization profiles. The single-vortex ansatz (SVA) describes a perfectly circular, field-independent in-plane vortex with a regularized out-of-plane core \cite{usov1993}; while it captures the topology, it does not by itself satisfy the magnetostatic boundary condition at the disk edge once the core is displaced. The two-vortex ansatz (TVA) corrects this by superposing SVA with an image vortex placed beyond the disk, generating the radial bending required to suppress edge surface charges \cite{guslienko2001}. Further along this hierarchy, the pole-free ansatz of Metlov and Guslienko enforces the absence of side surface charges in a more general way \cite{metlov2002}, and yields closed-form expressions for the magnetostatic energy and stiffness coefficients entering the Thiele formalism \cite{dewergifosse2023, khvalkovskiy2009}.

All members of this hierarchy share a structural limitation: each describes only how the equilibrium texture adapts to the disk geometry, with no internal degree of freedom for responding to external perturbations such as magnetic fields. Because projecting onto a rigid profile eliminates the degrees of freedom through which these perturbations enter, the resulting Thiele coefficients lack the field-induced corrections to stiffness, damping, and gyrotropic force that govern the core trajectory. Current practice circumvents this either through full micromagnetic simulations (accurate but computationally expensive, and giving little analytical insight) or through ad hoc corrections applied after projection, neither of which provides a systematic route to constructing the deformed magnetization profile. What is missing is not a better fixed profile, but an ansatz that is itself aware of the perturbation acting on it.

\begin{figure}[htbp]
\centering
\includegraphics[width=1\linewidth]{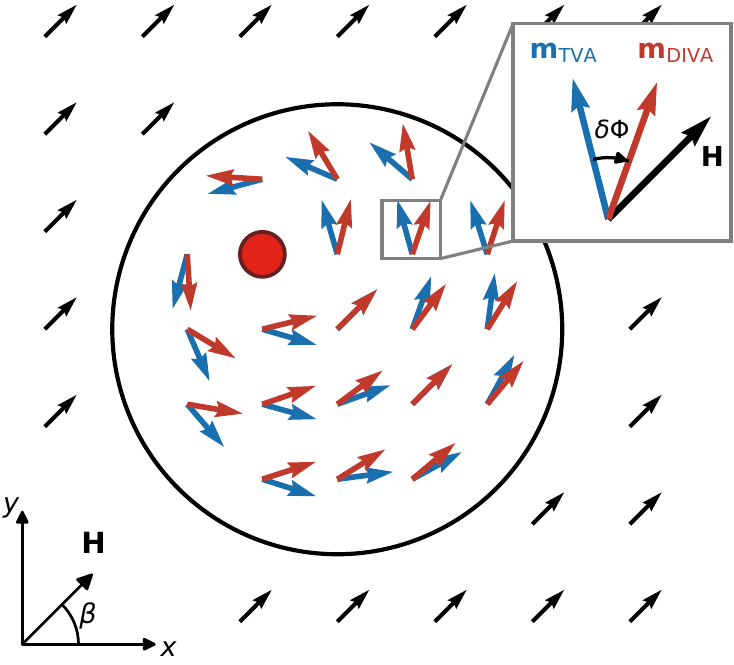}
\caption{Schematic of the DIVA construction for a magnetic vortex in a circular ferromagnetic nanodot under a uniform in-plane field. Black arrows indicate the in-plane field $\mathbf{H}$, reproduced in the coordinate reference (bottom left) at an angle $\beta$ from the $x$-axis. Inside the disk, the in-plane magnetization curls around the out-of-plane vortex core (red dot, here displaced from the disk center); two textures are overlaid on the same grid: the two-vortex ansatz (TVA, blue arrows), which represents only the displaced equilibrium vortex, and DIVA (red arrows), which additionally carries the field-induced deformation. The inset enlarges the boxed region and compares the local magnetization directions of the two ansätze, $\mathbf{m}_{\text{TVA}}$ (blue arrow) and $\mathbf{m}_{\text{DIVA}}$ (red arrow), against the local field direction (black arrow): the two ansätze differ by the field-induced angular deformation $\delta\Phi$, which tilts the magnetization toward $\mathbf{H}$.}
\label{fig:vortex}
\end{figure}

We address this gap by introducing the Deformed Image Vortex Ansatz (DIVA), a perturbation-aware ansatz in which the magnetization profile is generalized from an equilibrium texture to a deformed object that carries the response to an external perturbation as part of its definition. The deformation lives inside the profile that the Thiele projection acts upon, rather than being appended to the dynamics as a correction. The concept is not tied to any particular equilibrium profile: any ansatz from the existing hierarchy can serve as the base to which a perturbation-aware deformation is added, and the deformation itself may be derived analytically, inferred from data, or constructed by any route compatible with the symmetry and boundary structure of the problem.

We demonstrate the concept for a uniform, stationary in-plane field applied to a Permalloy nanodot (see Fig.~\ref{fig:vortex}), for which the deformation is analytically tractable. Symmetry-based analysis of the micromagnetic energy functional restricts the leading deformation to the $m=1$ harmonic, and a dominant-balance solution of the resulting Euler--Lagrange equation yields a closed-form interpolant for the radial profile. Benchmarked against full micromagnetic simulations on Permalloy disks of aspect ratio $t/R = 0.1$ and $0.0125$, this realization reduces the disk-averaged angular deviation by a factor of 3 to 6 relative to the two-vortex ansatz, depending on geometry and field, up to reduced core displacements $s \approx 0.8$.

\section{Methodology}
\label{sec:methodology}

The construction proceeds by two generic operations applied to the in-plane case: (i) symmetry-based selection of the deformation that the field drives; and (ii) solution of the Euler--Lagrange equation that follows from energy minimization.

\subsection{Physical setting and reference state}
\label{sec:reference_state}

We consider a ferromagnetic thin disk of radius $R$ and thickness $t$, with exchange stiffness $A$ and saturation magnetization $M_\text{s}$. In the thin-film regime ($t \ll R$), the magnetization is, to good approximation, uniform through the thickness, the standard quasi-2D assumption of vortex-state models~\cite{guslienko2002}, which reduces the problem to two dimensions in the disk plane. We use polar coordinates $(r,\theta)$ centered at the disk axis, with $r\in[0,R]$ the radial distance from the axis and $\theta\in[0,2\pi)$ the azimuthal angle measured from the $x$-axis; the associated orthonormal frame is $\{\hat{\mathbf{e}}_r,\hat{\mathbf{e}}_\theta,\hat{\mathbf{e}}_z\}$ with $\hat{\mathbf{e}}_r = (\cos\theta,\sin\theta,0)$ and $\hat{\mathbf{e}}_\theta = (-\sin\theta,\cos\theta,0)$. Outside the vortex core, we assume planar magnetization:
\begin{equation}
    \mathbf{m}(r,\theta) = \cos\Phi(r,\theta)\,\hat{\mathbf{e}}_x + \sin\Phi(r,\theta)\,\hat{\mathbf{e}}_y,
\end{equation}
where $\Phi$ is the in-plane angle of $\mathbf{m}$ measured from $\hat{\mathbf{e}}_x$, and the core is treated as a point singularity. The exchange length is $\ell_{\text{ex}} = \sqrt{2A/(\mu_0 M_\text{s}^2)}$.

Without any perturbation, the equilibrium is the ideal centered vortex:
\begin{equation}
\Phi_0(\theta) = \theta + C \frac{\pi}{2} \iff \mathbf{m}_0 = C\,\hat{\mathbf{e}}_\theta.
\end{equation}
This reference state carries no bulk magnetic charge and has zero Zeeman energy in any uniform field.

\subsection{Form of the field-induced deformation}

The full magnetization angle decomposes as:
\begin{equation}
\Phi(r,\theta, \mathbf{H}) = \Phi_0(\theta) + \delta\Phi(r,\theta, \mathbf{H}),
\end{equation}
where $\delta \Phi$ is the deformation angle induced by the external field. For a uniform in-plane field
\begin{equation}
    \mathbf{H} = H(\cos\beta,\, \sin\beta,\, 0),
\label{eq:applied_field}
\end{equation}
linearizing the Zeeman energy in $\delta\Phi$ produces a driving term proportional to $\sin(\Phi_0 - \beta)$, which is a pure $m=1$ Fourier mode in $\theta$. Higher harmonics are orthogonal to this driving term and receive no energy at first order in $H$ (see Sec.~S1 of the Supplemental Material~\cite{supplement}). The deformation ansatz therefore takes the form
\begin{equation}
\delta\Phi(r,\theta, \mathbf{H}) = f(r, H)\sin[\Phi_0(\theta) - \beta],
\label{eq:perturbative_ansatz}
\end{equation}
reducing the problem to the radial profile $f(r, H)$, which vanishes at the core, $f(r=0, H) = 0$, and with the field, $f(r, H=0)=0$, recovering the unperturbed vortex.

\subsection{Micromagnetic energy functional}

The full micromagnetic energy~\cite{hubert1998} encompasses exchange ($E_\text{ex}$), Zeeman ($E_Z$), magnetostatic ($E_{\text{ms}}$), magnetocrystalline anisotropy ($E_\text{A}$), magnetoelastic ($E_\text{me}$), and Dzyaloshinskii--Moriya interaction ($E_\text{DMI}$) terms. For Permalloy (Ni$_{80}$Fe$_{20}$), the cubic anisotropy constant $K_1$ is near-zero~\cite{yin2006} and the magnetostriction coefficient $\lambda_s \approx 0$~\cite{balakrishna2021}, making $E_\text{A}$ and $E_\text{me}$ negligible at room temperature. $E_\text{DMI}$ is absent in Permalloy due to its centrosymmetric crystal structure. The total energy therefore reduces to:
\begin{equation}
    E = E_{\text{ex}} + E_Z + E_{\text{ms}}.
\label{eq:total_energy}
\end{equation}

We substitute the perturbed angle $\Phi = \Phi_0 + f(r)\sin(\Phi_0-\beta)$ into each term and retain the $f$-dependent (deformation) contribution; the magnetization is taken uniform along the thickness, so the $z$-integral reduces to a factor $t$. Throughout, $\eta = r/R$ and $f' = df/d\eta$. The full derivations are given in Sec.~S2 of the Supplemental Material.

\paragraph{Exchange energy.}
The deformation raises the exchange energy~\cite{landau1935} by
\begin{equation}
\delta E_{\text{ex}}[f] = \pi At \int_0^1 \left[\eta\, f'(\eta)^2 + \frac{f(\eta)^2}{\eta}\right] d\eta.
\label{eq:dEex}
\end{equation}
The first term penalizes the radial gradient of $f$; the second is the intrinsic exchange cost of the $m=1$ curvature.

\paragraph{Zeeman energy.}
The ideal vortex carries zero Zeeman energy in any uniform in-plane field (see Sec.~\ref{sec:reference_state}); the field couples to the deformation only at first order in $f$~\cite{aharoni2000}:
\begin{equation}
\delta E_Z[f] = \pi\mu_0 M_\text{s} HtR^2 \int_0^1 f(\eta)\,\eta\,d\eta.
\label{eq:dEZ}
\end{equation}
This contribution is linear in $f$ and negative (stabilizing) for $f<0$: it is the driving term that produces the field-induced deformation.

\paragraph{Magnetostatic energy.}
The deformation generates bulk and surface magnetic charges linear in $f$, so the magnetostatic self-energy~\cite{hubert1998} is a nonlocal functional quadratic in $f$. The long-range dipolar coupling prevents an exact reduction of this nonlocal form to a local one, as already demonstrated in extended films~\cite{kalinikos1986}, and confinement in a nanodot does not lift this nonlocality. We therefore approximate it by a local single-mode functional. In plain terms, the true energy couples the charge at every point to the charge everywhere else; we replace it by an energy density that depends only on the local value of $f$, with a single coefficient, the in-plane demagnetizing factor $\mathcal{N}$, carrying the entire magnetostatic response of the $m=1$ harmonic:
\begin{equation}
\delta E_\text{ms}[f] \approx \mathcal{N}\,\pi\mu_0 M_\text{s}^2\, t R^2 \int_0^1 f(\eta)^2\,\eta\, d\eta,
\label{eq:Ems_local}
\end{equation}
where $\mathcal{N}$ depends only on the aspect ratio $t/R$. We approximate it by the in-plane demagnetizing factor of a uniformly magnetized cylinder of radius $R$ and thickness $t$, for which Sato and Ishii give the closed form~\cite{sato1989}
\begin{equation}
\mathcal{N} \approx \frac{t/R}{\sqrt{\pi} + 2\,t/R}.
\label{eq:N_sato}
\end{equation}
Equation~(\ref{eq:N_sato}) is an approximation rather than an identity, on two counts: the closed form itself deviates from the exact in-plane demagnetizing factor of a uniformly magnetized cylinder, and the vortex texture is nonuniform, which adds a smaller correction. A direct micromagnetic calibration of the $m=1$ demagnetizing factor (see Sec.~S4 of the Supplemental Material) shows that it underestimates $\mathcal{N}$ by ${\sim}12\%$ for $t/R = 0.1$ and by ${\sim}37\%$ for $t/R = 0.0125$, the deficit growing as the disk gets thinner. We nevertheless retain the closed form throughout, so that the construction stays analytical end to end, and quantify the resulting error against micromagnetic simulations in Sec.~\ref{sec:results}.

\subsection{Radial equation and scale separation}\label{sec:radial}

The second operation solves the Euler--Lagrange equation derived from the deformation energy. Minimizing $E[f] = E_0 + \delta E_\text{ex}[f] + \delta E_Z[f] + \delta E_\text{ms}[f]$ with respect to $f$ yields the governing equation for the radial profile (see Sec.~S3 of the Supplemental Material):
\begin{equation}
\eta^2 f''(\eta) + \eta f'(\eta) - \left(1 + \frac{\eta^2}{\mathcal{A}^2}\right)f(\eta) = \mathcal{B}\,\eta^2,
\label{eq:EL_equation}
\end{equation}
with the two dimensionless ratios
\begin{equation}
    \mathcal{A} = \frac{\ell_{\text{ex}}}{R\sqrt{2\mathcal{N}}}, \qquad \mathcal{B} = \frac{H}{H_{\text{ex}}},
\end{equation}
and $H_{\text{ex}} = 2A/(\mu_0 M_\text{s} R^2)$~\cite{aharoni2000}. Here $\mathcal{A} = \ell_\text{sat}/R$ compares the deformation saturation length $\ell_\text{sat} = \ell_\text{ex}/\sqrt{2\mathcal{N}}$ to the disk radius; for the geometries of Table~\ref{tab:params} it takes the values $\mathcal{A} \approx 0.16$ and $0.11$, so the small-parameter assumption holds only marginally and the scale separation carries a nominal $O(\mathcal{A}^2)$ error of a few percent. Meanwhile, $\mathcal{B}$ is the reduced field; since Eq.~(\ref{eq:EL_equation}) is linear with a driving term $\propto\mathcal{B}$, the profile scales linearly with the field, $f\propto\mathcal{B}$. Being second order in $\eta$, Eq.~(\ref{eq:EL_equation}) is closed by two boundary conditions, both fixed by physics rather than convenience: regularity at the core, $f(0)=0$, forced by finiteness of the exchange energy, and the natural free-edge condition $f'(1)=0$, which expresses the absence of any radial torque pinning the angle at the rim (see Sec.~S3 of the Supplemental Material).

Because $\mathcal{A}$ is small but not negligible (Table~\ref{tab:params}), Eq.~(\ref{eq:EL_equation}) separates into two scales: over the bulk of the disk the magnetostatic term dominates, giving a uniform plateau
\begin{equation}
f_\text{bulk} = -\frac{H}{2\mathcal{N} M_\text{s}},
\label{eq:f_outer}
\end{equation}
set by the Zeeman drive against the demagnetizing factor $\mathcal{N}$; the exchange term re-enters only in a core layer of width $\mathcal{A}$ that restores regularity, $f(0)=0$. Matching the plateau to this core layer motivates the closed-form interpolant (see Sec.~S3 of the Supplemental Material)
\begin{equation}
f(r) = -\frac{H}{2\mathcal{N}M_\text{s}}\!\left[1 - \exp\!\left(-\frac{r}{\ell_\text{sat}}\right)\right],
\label{eq:DIVA_profile_physical}
\end{equation}
which rises linearly from the core and saturates to $f_\text{bulk}$ over the length $\ell_\text{sat}$. It reproduces the plateau and the crossover scale exactly, but approximates the exact solution of Eq.~(\ref{eq:EL_equation}) rather than equalling it: the interpolant runs up to ${\sim}10\%$ of the plateau value too negative, the deviation peaking near $\eta \approx 0.2$ and decaying slowly outward (see Sec.~S3 of the Supplemental Material).

\subsection{DIVA profile}
\label{sec:DIVA_Profile}

The two operations above determine the deformation profile around a centered vortex, with the demagnetization factor $\mathcal{N}$ supplied in closed form by Eq.~(\ref{eq:N_sato}). Extending the result to a displaced core requires a composition rule with an edge-compatible equilibrium ansatz; the rule is specific to this realization, since it depends on which equilibrium ansatz is appropriate to the perturbation being described. For the in-plane case we use the two-vortex ansatz.

The preceding sections established that, for a centered vortex under a uniform external field, the in-plane angle deforms as $\Phi = \Phi_0 + f(r)\sin(\Phi_0 - \beta)$ (see Eq.~\ref{eq:perturbative_ansatz}), with the radial profile $f(r)$ given analytically by Eq.~(\ref{eq:DIVA_profile_physical}). This result was derived under the assumption of a centered vortex; a displaced vortex requires extending it to finite core positions.

To this end, we combine the deformation profile with the TVA~\cite{guslienko2001}. For a core at position $\mathbf{X}$, the DIVA in-plane angle is:
\begin{equation}
\begin{split}
\Phi_{\text{DIVA}}(\mathbf{r},\mathbf{X},\mathbf{H}) &=
\Phi^{\text{TVA}}(\mathbf{r},\mathbf{X}) \\
&\quad + f(|\mathbf{r}-\mathbf{X}|, H) \\
&\quad \times \sin\!\left[\Phi^{\text{TVA}}(\mathbf{r},\mathbf{X}) - \beta\right],
\end{split}
\label{eq:DIVA_displaced}
\end{equation}
where the TVA term provides the displaced, edge-compatible in-plane texture, and the perturbation adds the field-induced $m=1$ deformation evaluated at the distance from the displaced core. This rigid translation of the deformation onto the displaced core is permitted precisely by the shape of $f$: the profile is constant (the plateau $f_\text{bulk}$) throughout the bulk and varies only within the core layer of width $\ell_\text{sat}$ centered on the core. The deformation is thus effectively pinned to the core, so recentering it on $\mathbf{X}$ leaves the bulk texture unchanged and merely carries the localized core layer along with the core. $\Phi_\text{DIVA}$ encodes only the in-plane angle; the complete 3D magnetization additionally requires an out-of-plane core profile (see Sec.~\ref{sec:results}).

Although $f(r)$ was derived for a centered vortex, its validity at finite displacement is set by the localization scale $\mathcal{A} = \ell_\text{sat}/R$, not by the displacement $s = |\mathbf{X}|/R$ directly. Because $f$ reaches its plateau within $\sim\ell_\text{sat} = \mathcal{A}R$ of the core, the centered-vortex profile carries over to a displaced core as long as this core layer remains inside the disk,
\begin{equation}
1 - s \;\gtrsim\; \mathcal{A}.
\label{eq:validity_window}
\end{equation}
Within this window, $\Phi_\text{DIVA}$ is a fully analytical in-plane texture valid up to large core displacements, requiring no micromagnetic relaxation.

\section{Results and Discussion}
\label{sec:results}

We compare DIVA against the two-vortex ansatz (TVA) and full micromagnetic simulations (MMS) performed with MuMax+~\cite{mumaxplus}. Two disk geometries spanning an eightfold range of aspect ratio $t/R$ are considered: $t/R = 0.1$ ($R = \SI{100}{\nano\metre}$, $t = \SI{10}{\nano\metre}$) and $t/R = 0.0125$ ($R = \SI{400}{\nano\metre}$, $t = \SI{5}{\nano\metre}$). Material parameters are those of Permalloy ($M_\text{s} = \SI{8e5}{\ampere\per\metre}$, $A = \SI{1.07e-11}{\joule\per\metre}$), the chirality is fixed to $C = +1$, and the in-plane field is applied along $\beta = 0$. The in-plane demagnetizing factor $\mathcal{N}$ is evaluated per geometry from Eq.~(\ref{eq:N_sato}), giving $\mathcal{N} = 0.051$ for $t/R = 0.1$ and $\mathcal{N} = 0.0070$ for $t/R = 0.0125$; these are the values behind every DIVA curve reported below, and the calibrated values of Sec.~S4 are used nowhere in the benchmarks. Table~\ref{tab:params} collects the derived length scales for both geometries.

\begin{table}[t]
\caption{Derived parameters for the two benchmark geometries (Permalloy, $\ell_\text{ex} \approx \SI{5.2}{\nano\metre}$). $\mathcal{N}$ is the Sato--Ishii value of Eq.~(\ref{eq:N_sato}); the last column is the reduced-displacement window $1 - s \gtrsim \mathcal{A}$, Eq.~(\ref{eq:validity_window}).}
\label{tab:params}
\begin{ruledtabular}
\begin{tabular}{ccccccc}
$t/R$ & $R$ (\si{\nano\metre}) & $t$ (\si{\nano\metre}) & $\mathcal{N}$ & $\ell_\text{sat}$ (\si{\nano\metre}) & $\mathcal{A}$ & $1-\mathcal{A}$\\
\hline
0.1 & 100 & 10 & 0.051 & 16 & 0.16 & 0.84\\
0.0125 & 400 & 5 & 0.0070 & 44 & 0.11 & 0.89\\
\end{tabular}
\end{ruledtabular}
\end{table}

The comparison is performed at fixed core position: for each field value, the equilibrium core position $\mathbf{X}$ is obtained from MMS by relaxing the magnetization, and TVA and DIVA are then evaluated analytically at the same $\mathbf{X}$ (with reduced displacement $s = |\mathbf{X}|/R$). This isolates the quality of the in-plane texture from any error in core-position prediction. Both TVA and DIVA treat the vortex core as a point singularity in the in-plane angle; for the total-energy comparison of Sec.~\ref{sec:energy}, the singularity is regularized with the Feldtkeller out-of-plane core profile~\cite{feldtkeller1965} of radius $r_c = \sqrt{2}\,\ell_\text{ex}\,(1 + 0.39\,t/\ell_\text{ex})^{1/3}$~\cite{kravchuk2006}, applied identically to both ansätze so that they differ only through their in-plane textures.

\subsection{Validation of the deformation profile}
\label{sec:validation}

The DIVA construction relies on two assumptions: (i) the field-induced
deformation couples exclusively to the $m = 1$ harmonic, with the
angular dependence $\sin[\Phi(\theta) - \beta]$, and (ii) the radial profile
$f(r)$ derived from the dominant-balance analysis captures the
exchange-to-magnetostatic crossover. Both assumptions can be tested directly
from the spatial map of the angular residual
$\Delta\Phi(\mathbf{r}) = \Phi_\text{MMS}(\mathbf{r}) - \Phi_\text{ansatz}(\mathbf{r})$
shown in Figs.~\ref{fig:anglemapR100h10}
and~\ref{fig:anglemapR400h5}.

For TVA, $\Delta\Phi$ exhibits a clear two-lobe pattern aligned
perpendicular to $\mathbf{H}$, with positive and negative lobes of
comparable amplitude on either side of the disk, peaking at
approximately \ang{22} of error in the $t/R = 0.1$ disk and \ang{24} in
the $t/R = 0.0125$ disk. This signature is
that of the missing $m=1$ component, and its
angular orientation matches the $\sin[\Phi(\theta) - \beta]$
dependence postulated in Eq.~(\ref{eq:perturbative_ansatz}). When
the same map is computed using DIVA instead, this two-lobe
pattern is strongly reduced: in the $t/R = 0.1$ disk the residual over
the disk body drops to approximately \ang{5} (a reduction of roughly a
factor of four), with a weak two-lobe residual reaching \ang{9} in a
narrow band at the disk boundary. In the $t/R = 0.0125$ disk the
suppression is also clear, but a two-lobe residual of larger amplitude,
up to \ang{10}, persists across the disk body.

In the $t/R = 0.1$ disk, the near-complete suppression of the two-lobe
pattern confirms that the analytical $f(r)$ captures the bulk of the
$m=1$ deformation. In the $t/R = 0.0125$ disk it captures most of it, but a
residual of the same two-lobe ($m=1$) character persists across the
disk body, indicating that $f(r)$ underestimates the deformation
amplitude in this geometry.
The symmetry argument behind Eq.~(\ref{eq:perturbative_ansatz}) is tested most directly by the TVA residual, which measures the deformation itself: in both geometries it is dominated by the single two-lobe ($m=1$) pattern that argument predicts. The residual left by DIVA is several times smaller, and we do not characterize its harmonic content here.

\subsection{Mean angular error}
\label{sec:meanerror}

We now turn to aggregate accuracy of the in-plane realization. A quantitative measure is provided by the disk-averaged absolute angular error,
\begin{equation}
\langle |\Delta\Phi| \rangle =
\frac{1}{\mathcal{S}} \int |\Delta\Phi(\mathbf{r})|\, d^2r,
\label{eq:meanerror}
\end{equation}
where $\mathcal{S}$ is the area of the integration domain.

\begin{figure}[t]
\centering
\includegraphics[width=\linewidth]{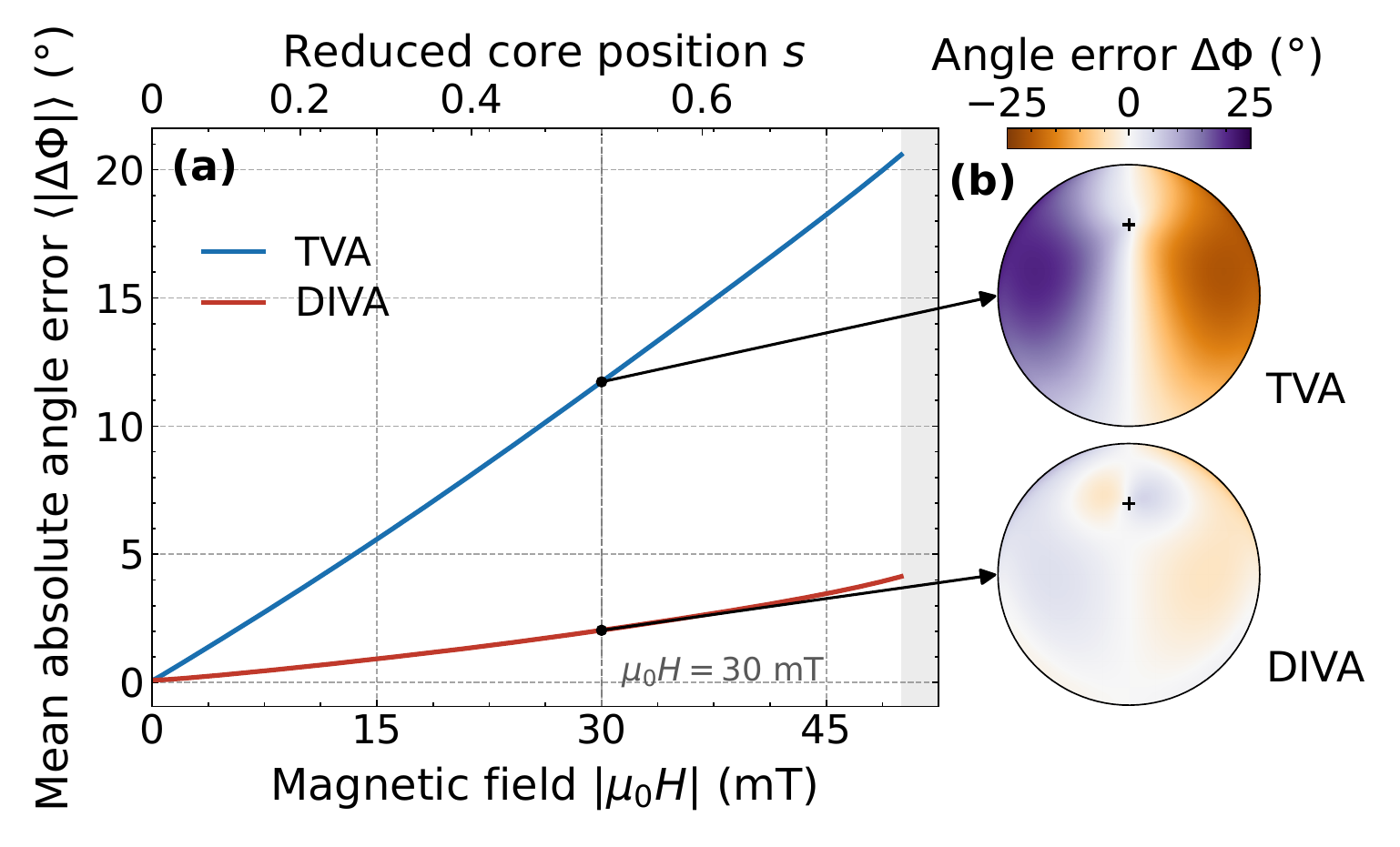}
\caption{Angular accuracy of the two ansätze for the $t/R = 0.1$ disk ($R = \SI{100}{\nano\metre}$, $t = \SI{10}{\nano\metre}$, Permalloy parameters, $\beta=0$, $C=+1$).
    (a) Mean absolute angle error $\langle |\Delta\Phi| \rangle$ between the analytical ansätze and the relaxed MuMax+ configuration, as a function of the applied field $|\mu_0 H|$ (bottom axis) and of the corresponding reduced core position $s=|\mathbf{X}|/R$ (top axis). Blue line: two-vortex ansatz (TVA). Red line: deformed image vortex ansatz (DIVA). Curves are plotted up to vortex-core expulsion; the shaded band marks $s > 1-\mathcal{A}$, beyond the validity window of Eq.~(\ref{eq:validity_window}) (Table~\ref{tab:params}): for this geometry the core is expelled at $s\approx0.8$, before the window is reached, so no data enter the shaded region.
    (b) Spatial maps of the signed angular residual $\Delta\Phi=\Phi_\text{MMS} - \Phi_{\text{ansatz}}$ for TVA (top) and DIVA (bottom), evaluated at the operating point marked in (a) by the filled circle on each curve and by the vertical dashed line, $\mu_0 H = \SI{30}{\milli\tesla}$ ($s \approx 0.51$); the cross marks the vortex-core position. Both maps share the color scale shown above them, which spans $\pm\ang{25}$. The TVA residual exhibits the two-lobe pattern characteristic of the missing $m=1$ deformation; in DIVA it is strongly suppressed.
    }
\label{fig:anglemapR100h10}
\end{figure}

Figure~\ref{fig:anglemapR100h10} reports $\langle |\Delta\Phi| \rangle$ for the $t/R = 0.1$ disk as a function of the external field, with the corresponding reduced core displacement on the top axis. The TVA error grows monotonically to ${\sim}\ang{21}$ at $\mu_0 H = \SI{50}{\milli\tesla}$ (vortex core expulsion at $s \approx 0.8$), while DIVA reaches only ${\sim}\ang{4}$ at the same field.

\begin{figure}[t]
\centering
\includegraphics[width=\linewidth]{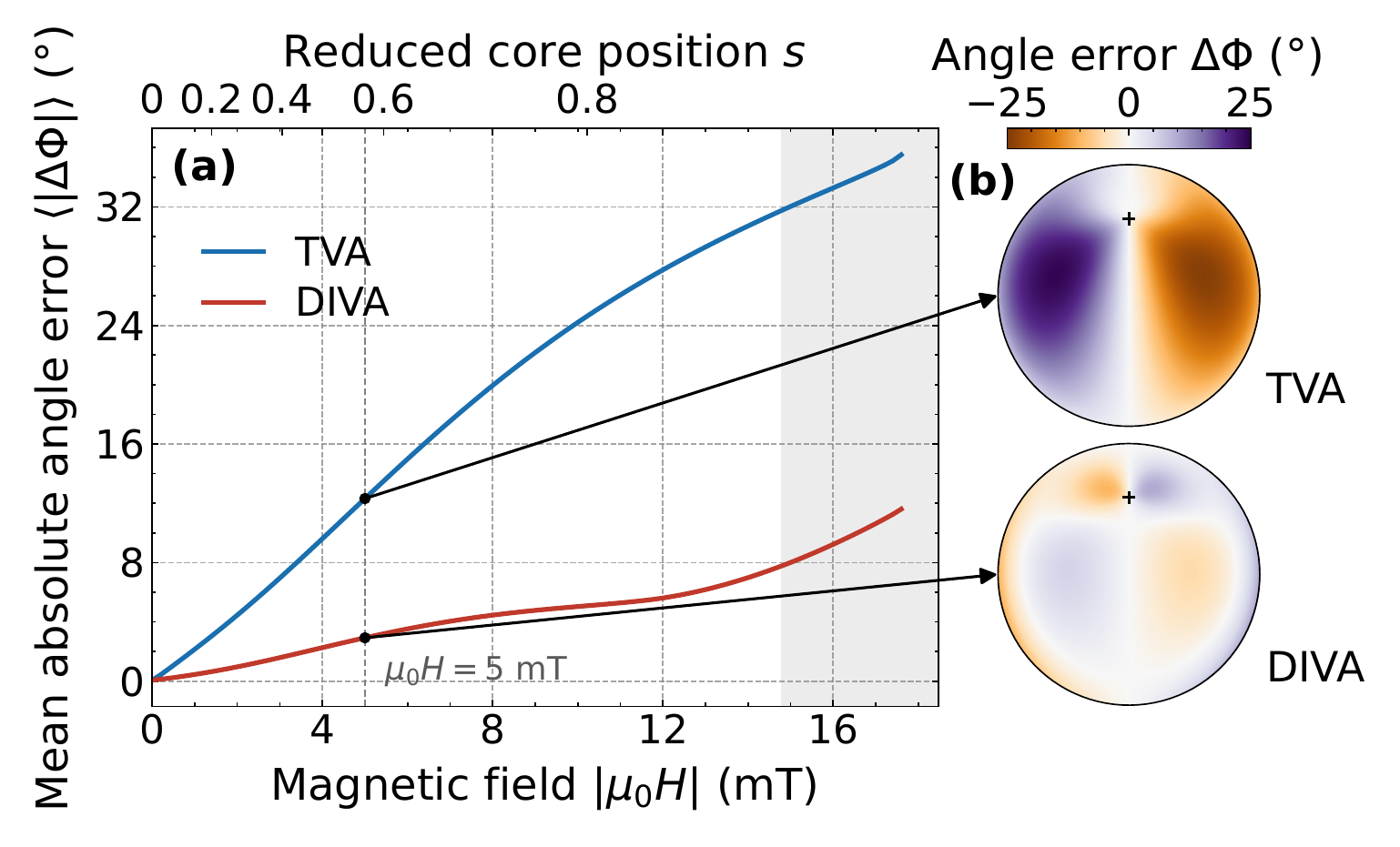}
\caption{Angular accuracy of the two ansätze for the $t/R = 0.0125$ disk ($R = \SI{400}{\nano\metre}$, $t = \SI{5}{\nano\metre}$, Permalloy parameters, $\beta=0$, $C=+1$).
    (a) Mean absolute angle error $\langle |\Delta\Phi| \rangle$ between the analytical ansätze and the relaxed MuMax+ configuration, as a function of the applied field $|\mu_0 H|$ (bottom axis) and of the corresponding reduced core position $s=|\mathbf{X}|/R$ (top axis). Blue line: two-vortex ansatz (TVA). Red line: deformed image vortex ansatz (DIVA). Curves are plotted up to vortex-core expulsion; the shaded band marks $s > 1-\mathcal{A}$, beyond the validity window of Eq.~(\ref{eq:validity_window}); unlike the $t/R=0.1$ disk, here the core survives into this region before being expelled near the end of the sweep.
    (b) Spatial maps of the signed angular residual $\Delta\Phi=\Phi_\text{MMS} - \Phi_{\text{ansatz}}$ for TVA (top) and DIVA (bottom), evaluated at the operating point marked in (a) by the filled circle on each curve and by the vertical dashed line, $\mu_0 H = \SI{5}{\milli\tesla}$ ($s \approx 0.57$); the cross marks the vortex-core position. Both maps share the color scale shown above them, which spans $\pm\ang{25}$, the same range as in Fig.~\ref{fig:anglemapR100h10}. The TVA two-lobe pattern is strongly reduced in DIVA, but a weaker residual persists across the disk body and along the rim.
    }
\label{fig:anglemapR400h5}
\end{figure}

Figure~\ref{fig:anglemapR400h5} repeats the comparison for the $t/R = 0.0125$ disk, where the larger lateral extent enhances magnetostatic effects and lowers the field required to displace the core. TVA grows monotonically to ${\sim}\ang{35}$ before $s = 1$; DIVA stays far below it, but its growth is not uniform: the error flattens into a shoulder around $s \approx 0.8$ and then climbs again, more steeply than at any smaller displacement, to ${\sim}\ang{11}$ at the end of the sweep.

The two geometries therefore display qualitatively different DIVA behavior: steady growth in the $t/R = 0.1$ disk, and the nonmonotonic slope just described in the $t/R = 0.0125$ disk. We report this shoulder as an observation; its origin is not established here, and resolving it would require sweeping a third aspect ratio to test whether its position tracks $\mathcal{A}$. DIVA stays well below TVA in both geometries throughout.

\subsection{Total energy}
\label{sec:energy}

\begin{figure}[t]
\centering
\includegraphics[width=\linewidth]{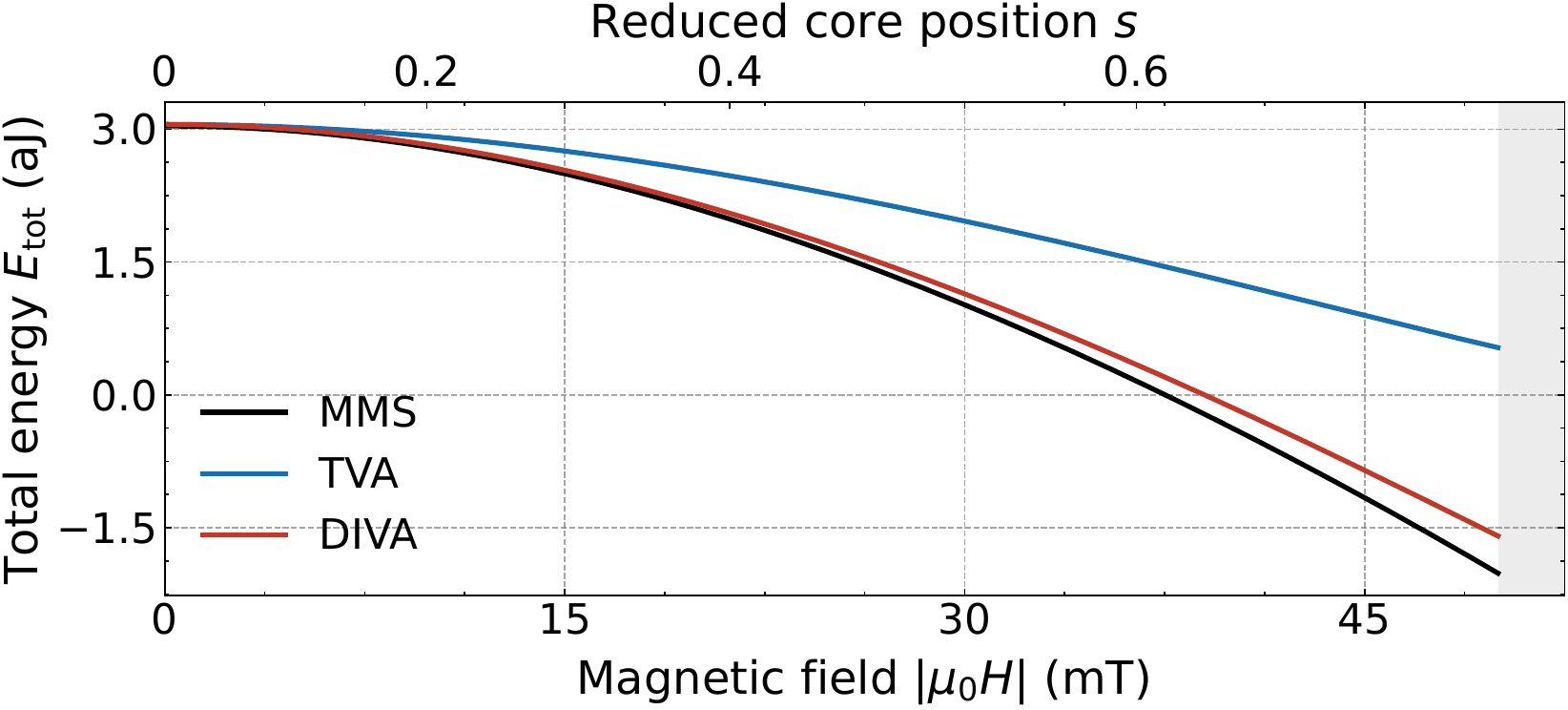}
\caption{Total micromagnetic energy $E_\text{tot}$ of the displaced vortex as a function of the applied field $|\mu_0 H|$ (bottom axis) and of the associated reduced core position $s$ (top axis), for the $t/R = 0.1$ disk ($R = \SI{100}{\nano\metre}$, $t = \SI{10}{\nano\metre}$, Permalloy parameters, $\beta=0$, $C=+1$). Black: full micromagnetic simulation (MMS). Blue: TVA evaluated at the MMS core position. Red: DIVA evaluated at the same position. Both ansätze carry the same Feldtkeller out-of-plane core, so they differ only through their in-plane textures. The shaded band marks $s > 1-\mathcal{A}$, beyond the validity window of Eq.~(\ref{eq:validity_window}); as in Fig.~\ref{fig:anglemapR100h10}, the core is expelled at $s\approx0.8$, before this window, so the shaded region contains no data.
    }
\label{fig:energyR100h10}
\end{figure}

A complementary probe of the in-plane realization is the total micromagnetic
energy itself. Whereas the angular residual measures the pointwise fidelity of the magnetization profile, the total
energy is a global functional that combines a linear Zeeman gain
with quadratic exchange and magnetostatic costs in $f$; it therefore
weights the field-induced deformation differently from the pointwise
angular measure of Sec.~\ref{sec:meanerror}. We restrict this comparison
to the $t/R = 0.1$ disk ($R = \SI{100}{\nano\metre}$, $t = \SI{10}{\nano\metre}$); the
analogous comparison for the $t/R = 0.0125$ disk ($R = \SI{400}{\nano\metre}$,
$t = \SI{5}{\nano\metre}$) shows the same behavior of the total energy and is
reported in Sec.~S5.2 of the Supplemental Material, where the individual
contributions are seen to deviate in the opposite sense.
Figure~\ref{fig:energyR100h10} shows the total micromagnetic energy
$E_\text{tot}$ as a function of the external field for this geometry,
computed by direct evaluation of the energy functional on the relaxed
MMS configuration (black), on the TVA profile at the MMS core
position (blue), and on the DIVA profile at the same position (red).

At zero field the TVA, DIVA, and MMS curves agree to
\SI{0.02}{\atto\joule} out of ${\sim}\SI{3.0}{\atto\joule}$: both ansätze
reduce to the centered ideal vortex with the Feldtkeller
core~\cite{feldtkeller1965} when $H = 0$, and this core reproduces the
energy of the fully relaxed configuration resolved by the magnetostatic
solver to within that offset. The residual is the cost of the core
regularization, which is localized on the scale $\ell_\text{ex}$ and
decouples from the bulk deformation at the perturbative order considered
here; being field-independent, it leaves any field-dependent deviation
that follows as a property of the in-plane texture alone.

As the field increases, the MMS energy decreases nonlinearly, driven
by the Zeeman gain associated with the field-induced deformation and
the displacement of the core. The TVA energy follows the same
qualitative trend but accumulates a growing deviation relative
to MMS: the TVA--MMS gap widens from zero at $H = 0$ to
${\sim}\SI{2.5}{\atto\joule}$ at $\mu_0 H = \SI{50}{\milli\tesla}$, growing
approximately quadratically in $H$. This excess is the expected
signature of an ansatz evaluated at the correct displaced core
position but lacking the field-induced $m=1$ deformation. Because the
deformation amplitude grows linearly with field ($f \propto H$,
Sec.~\ref{sec:radial}), the energy released by relaxing into it is
second order, $\propto H^2$; an ansatz that omits the deformation
therefore overestimates the energy by an amount growing quadratically
with field, consistent with the observed gap.

DIVA, by contrast, tracks the MMS energy across the entire field
range: the DIVA--MMS gap grows only to ${\sim}\SI{0.4}{\atto\joule}$ at
$\mu_0 H = \SI{50}{\milli\tesla}$, roughly six times smaller than
the TVA deviation at the same field. This small residual is consistent
with higher harmonics ($m \geq 2$) excited by the displaced
core and the deformation profile near the disk edge, neither of which
is retained at the leading order of the construction. These observations demonstrate
that DIVA captures the field dependence of the bulk deformation energy
at fixed core position with an accuracy approaching that of the
underlying micromagnetic solver, at negligible computational cost.

\section{Conclusion}

Rather than a magnetization profile selected to match the equilibrium geometry of the disk, DIVA is an ansatz that carries the response to an external perturbation as part of its definition: the deformation lives inside the magnetization profile that the Thiele projection acts upon, rather than being appended to the dynamics as a correction. The concept is method-agnostic: the deformation can be derived analytically, inferred from data, or constructed by any route compatible with the symmetry and boundary structure of the problem. In this sense DIVA complements the equilibrium-ansatz hierarchy without competing with it, adding a perturbation-aware companion to profiles that describe only the equilibrium adaptation of the texture to its geometry.

For the case of a uniform, stationary in-plane field applied to a Permalloy nanodot, the deformation is constructed analytically through two operations: symmetry-based selection of the $m=1$ harmonic and dominant-balance solution of the resulting Euler--Lagrange equation. Composition with the two-vortex ansatz extends the profile to displaced core configurations. Benchmarked against MuMax+ on Permalloy disks of aspect ratio $t/R = 0.1$ and $0.0125$, this realization reduces the disk-averaged angular deviation by a factor of 3 to 6 relative to the two-vortex ansatz, depending on geometry and field, and, for the $t/R = 0.1$ disk, tracks the total micromagnetic energy of the relaxed configuration to within ${\sim}\SI{0.4}{\atto\joule}$ at the largest field considered (roughly six times closer than the two-vortex ansatz), with the $t/R = 0.0125$ disk behaving the same way in the total energy, though through per-contribution deviations of the opposite sign (see Sec.~S5.2 of the Supplemental Material).

This realization has several limitations, in the order the derivation introduces them. The analysis is carried to first order in the field amplitude and retains only the $m=1$ harmonic, so harmonics $m \geq 2$ excited by finite core displacement are not described and contribute to the residual observed in the benchmarks of Sec.~\ref{sec:results}. The magnetostatic energy is reduced from a nonlocal functional to a local single-mode one characterized by a single demagnetizing factor $\mathcal{N}$ (Eq.~\ref{eq:Ems_local}), a reduction that cannot be made exact for a nanodot. That factor is then taken from the uniform-cylinder estimate of Eq.~(\ref{eq:N_sato}) rather than from the direct micromagnetic calibration of Sec.~S4 of the Supplemental Material. The radial profile of Eq.~(\ref{eq:DIVA_profile_physical}) is an interpolant reproducing both asymptotes and the crossover scale, not the exact solution of Eq.~(\ref{eq:EL_equation}). Transplanting that centered-vortex profile onto a displaced core is justified only while the core layer remains inside the disk, Eq.~(\ref{eq:validity_window}). Finally, the out-of-plane core profile is treated as a separable Feldtkeller regularization whose energy contribution is field-independent at the order considered and decouples from the deformation theory.

Beyond the in-plane field treated here, the concept is expected to apply to other perturbations whose source admits a clear symmetry: out-of-plane fields and Oersted contributions associated with the driving current in spin-torque oscillators each populate a different harmonic. The most immediate application target is the construction of perturbation-aware Thiele coefficients in which the deformation modifies the stiffness, damping, and gyrotropic terms governing core dynamics. Systematic inclusion of higher harmonics would extend the construction toward the finite-displacement regime relevant to large-amplitude oscillations.

The in-plane case shows that perturbation-aware ansätze are both constructible and quantitatively accurate; whether the same accuracy holds for other perturbations depends on each admitting its own tractable solution.

\begin{acknowledgments}
Computational resources have been provided by the Consortium des Équipements de Calcul Intensif (CÉCI), funded by the Fonds de la Recherche Scientifique de Belgique (F.R.S.-FNRS) under Grant No. 2.5020.11 and by the Walloon Region.
\end{acknowledgments}

\section*{Competing Interests}

The authors have no conflicts to disclose.

\section*{Author Contributions}

\textbf{T.G.C.}: conceptualization, methodology, software,
investigation, visualization, writing.
\textbf{C.D.}: software, investigation, reviewing.
\textbf{S.d.W.}: methodology, reviewing.
\textbf{F.A.A.}: conceptualization, methodology, supervision, funding
acquisition, reviewing and editing.

\section*{Funding}

T.G.C. is funded by Sirris and F.A.A. is a Research Associate of the F.R.S.-FNRS. The funders had no role in study design, data collection, data analysis, interpretation, or manuscript writing.

\section*{Data Availability}

Data generated during the current study are available
from the corresponding author on reasonable request.

%

\clearpage
\onecolumngrid
\startsupplement

\begin{center}
  {\large\bfseries Supplemental Material}\\[0.6em]
  {\large The Deformed Image Vortex Ansatz: A Perturbation-Aware
   Description of Magnetic Vortices in In-Plane Fields}\\[0.8em]
  Thomas G. Coppée, Colin Ducarme, Simon de Wergifosse,
  and Flavio Abreu Araujo
\end{center}
\vspace{1.5em}

\section{Symmetry-based selection of the $m=1$ harmonic}
\label{sec:supp_channel}

This section shows why a uniform in-plane field deforms the vortex only in
the $m=1$ harmonic (a single oscillation around the disk) and why
the deformation takes the fixed angular shape $\sin(\Phi_0-\beta)$. We write
the in-plane angle as $\Phi=\Phi_0+\delta\Phi$, with the ideal vortex
$\Phi_0(\theta)=\theta+C\pi/2$ and a small field-induced deformation
$\delta\Phi$, using the planar magnetization
$\mathbf{m}=(\cos\Phi,\sin\Phi,0)$ and the field
$\mathbf{H}=H(\cos\beta,\sin\beta,0)$.

Of the three energies, only the Zeeman energy depends on the field, so only
it can drive a deformation. With $\mathbf{H}\cdot\mathbf{m}=H\cos(\Phi-\beta)$,
expanding to first order in $\delta\Phi$ gives the Zeeman density
\begin{equation}
-\mu_0 M_\text{s}\,\mathbf{H}\cdot\mathbf{m}
= -\mu_0 M_\text{s} H\cos(\Phi_0-\beta)
\;\underbrace{+\;\mu_0 M_\text{s} H\,\delta\Phi\,\sin(\Phi_0-\beta)}_{\text{drive on }\delta\Phi}
\;+\;O(\delta\Phi^2).
\label{eq:supp_zeeman_drive}
\end{equation}
The first term is the Zeeman energy of the ideal vortex, which
integrates to zero over the disk; the second is the drive felt by the
deformation. Its angular factor
$\sin(\Phi_0-\beta)=\sin(\theta+C\pi/2-\beta)$ completes a single
oscillation around the disk: a pure $m=1$ pattern, with no overlap on any
other harmonic, $\int_0^{2\pi} e^{im\theta}\sin(\theta+C\pi/2-\beta)\,d\theta=0$
for $m\neq\pm1$.

The exchange and magnetostatic energies act in the opposite way. At zero
field the ideal vortex already minimizes both: it produces no magnetic
charges (its magnetization is divergence-free in the bulk and tangent to the
edge~\cite{sm:usov1993,sm:guslienko2001}) and sits at an exchange minimum. Neither
energy therefore has a term linear in $\delta\Phi$; a deformation only
raises them, at quadratic order. The field thus lowers the energy
solely through the $m=1$ part of $\delta\Phi$, while any other angular
harmonic would cost exchange and magnetostatic energy with nothing to gain
from the drive. Since the disk is rotationally symmetric the different
harmonics do not mix, so minimizing the energy keeps the $m=1$ component and
sets all others to zero. The deformation then inherits the exact angular
shape of the drive, leaving only a radial profile $f$ to be determined:
\begin{equation}
\delta\Phi(r,\theta,H) = f(r,H)\,\sin\!\big[\Phi_0(\theta)-\beta\big].
\end{equation}
Because the drive is proportional to $H$ while the restoring energies are
not, $f$ is linear in the field, $f\propto H$. Harmonics $|m|\ge2$ appear
only at second order, $O(H^2)$, through the displaced core and the quadratic
terms neglected in Eq.~(\ref{eq:supp_zeeman_drive}).

\section{Deformation contributions to the micromagnetic energy functional}
\label{sec:supp_energy}

This section derives the deformation contributions to the exchange,
Zeeman and magnetostatic energies quoted as Eqs.~(8)--(10) of the main
text. We work
with the perturbed in-plane angle
\begin{equation}
\Phi(r,\theta) = \Phi_0(\theta) + f(r)\sin\!\left[\Phi_0(\theta)-\beta\right],
\qquad \Phi_0(\theta) = \theta + C\tfrac{\pi}{2},
\label{eq:supp_phi}
\end{equation}
so that $\partial_r\Phi_0 = 0$ and $\partial_\theta\Phi_0 = 1$. The
magnetization is taken uniform along the thickness, reducing every
$z$-integral to a factor $t$, and we use the reduced radius
$\eta = r/R$ with $f' = df/d\eta$. We repeatedly use the azimuthal
averages
\begin{equation}
\int_0^{2\pi}\!\!\sin^2(\Phi_0-\beta)\,d\theta
= \int_0^{2\pi}\!\!\cos^2(\Phi_0-\beta)\,d\theta = \pi,
\quad
\int_0^{2\pi}\!\!\cos(\Phi_0-\beta)\,d\theta = 0.
\label{eq:supp_averages}
\end{equation}

\subsection{Exchange energy}

The exchange energy is~\cite{sm:landau1935}
\begin{equation}
E_\text{ex} = A\int_V (\nabla\Phi)^2\,dV
= At\int_0^R\!\!\int_0^{2\pi}
\left[(\partial_r\Phi)^2 + \frac{(\partial_\theta\Phi)^2}{r^2}\right]r\,dr\,d\theta.
\label{eq:supp_Eex_polar}
\end{equation}
Inserting Eq.~(\ref{eq:supp_phi}), the gradients decompose as
$\partial_r\Phi = f'(r)\sin(\Phi_0-\beta)$ and
$\partial_\theta\Phi = 1 + f(r)\cos(\Phi_0-\beta)$. Integrating over
$\theta$ with Eq.~(\ref{eq:supp_averages}), the zeroth-order term
$2\pi/r^2$ is constant in $f$ and is discarded, while the linear-in-$f$
term vanishes because $\int_0^{2\pi}\cos(\Phi_0-\beta)\,d\theta = 0$.
The surviving deformation contribution is
\begin{equation}
\delta E_{\text{ex}}[f] = \pi At \int_0^1 \left[\eta\, f'(\eta)^2 + \frac{f(\eta)^2}{\eta}\right] d\eta.
\label{eq:supp_dEex}
\end{equation}

\subsection{Zeeman energy}

The Zeeman energy is~\cite{sm:aharoni2000}
\begin{equation}
E_Z = -\mu_0 M_\text{s} \int_V \mathbf{H}\cdot\mathbf{m}\,dV
= -\mu_0 M_\text{s} H t \int_0^R\!\!\int_0^{2\pi} \cos(\Phi-\beta)\,r\,dr\,d\theta,
\label{eq:supp_EZ}
\end{equation}
where we used $\mathbf{H}\cdot\mathbf{m} = H\cos(\Phi-\beta)$ for the in-plane
field $\mathbf{H} = H(\cos\beta,\sin\beta,0)$ and planar magnetization
$\mathbf{m} = (\cos\Phi,\sin\Phi,0)$. The zeroth-order term vanishes
identically since the ideal vortex carries zero Zeeman energy in any
uniform in-plane field. Expanding to first order,
$\cos(\Phi-\beta) \approx \cos(\Phi_0-\beta) - f(r)\sin^2(\Phi_0-\beta)$,
and integrating over $\theta$ with Eq.~(\ref{eq:supp_averages}) gives
\begin{equation}
\delta E_Z[f] = \pi\mu_0 M_\text{s} HtR^2 \int_0^1 f(\eta)\,\eta\,d\eta.
\label{eq:supp_dEZ}
\end{equation}

\subsection{Magnetostatic energy}

The magnetostatic self-energy may be written compactly as the Coulomb
self-energy of the total magnetic charge density~\cite{sm:hubert1998},
\begin{equation}
E_\mathrm{ms}
=
\frac{\mu_0}{8\pi}
\iint
\frac{\rho_m(\mathbf r)\rho_m(\mathbf r')}
{|\mathbf r-\mathbf r'|}
\,d^3r\,d^3r',
\label{eq:supp_Ems_general}
\end{equation}
where
\[
\rho_m=-\nabla\!\cdot\!\mathbf M
\]
include both the volume
charge $-M_\text{s}\nabla\!\cdot\!\mathbf m$ and the surface charge
$M_\text{s}\mathbf m\!\cdot\!\hat{\mathbf n}$.

Writing
\[
\mathbf m=\mathbf m_0+\delta\mathbf m,
\]
where $\delta\mathbf m$ is generated by the angular perturbation
$\delta\Phi=f(r)\sin(\Phi_0-\beta)$, the charge density expands as
\[
\rho_m=\rho_m^{(0)}+\delta\rho_m.
\]
For the ideal vortex,
$\mathbf m_0=C\hat{\mathbf e}_\theta$,
one has
$\rho_m^{(0)}=0$:
the bulk charge vanishes because
$\nabla\!\cdot\!\mathbf m_0=0$,
while the edge charge vanishes because the magnetization is everywhere
tangential to the disk boundary~\cite{sm:usov1993,sm:guslienko2001}.
Consequently,
\[
\rho_m=\delta\rho_m,
\]
and the magnetostatic correction is simply
\begin{equation}
\delta E_\mathrm{ms}[f]
=
\frac{\mu_0}{8\pi}
\iint
\frac{\delta\rho_m(\mathbf r)\,
      \delta\rho_m(\mathbf r')}
     {|\mathbf r-\mathbf r'|}
\,d^3r\,d^3r'.
\label{eq:supp_Ems_expansion}
\end{equation}

The perturbation rotates the azimuthal magnetization toward the radial
direction,
\[
\delta\mathbf m
=
-C\,\delta\Phi\,\hat{\mathbf e}_r
=
-C f(r)\sin(\Phi_0-\beta)\,\hat{\mathbf e}_r,
\]
so the induced charge density $\delta\rho_m$ is linear in $f$ and
Eq.~(\ref{eq:supp_Ems_expansion}) is a nonlocal quadratic functional
of the radial profile $f$.

Three properties of $\delta E_\mathrm{ms}[f]$ follow immediately from
Eq.~(\ref{eq:supp_Ems_expansion}). It is quadratic in $f$, because the
induced charge is linear in $f$; it is positive, being the
self-energy of a real charge distribution; and it scales as
$\mu_0M_\text{s}^2$ times the disk volume. We approximate it by the
lowest-order local quadratic functional,
\begin{equation}
\delta E_\mathrm{ms}[f]
\approx
\mathcal N\mu_0M_\text{s}^2
\int_V
|\delta\mathbf m|^2\,dV,
\label{eq:supp_Ems_ansatz}
\end{equation}
where the dimensionless coefficient $\mathcal N>0$ depends only on the
aspect ratio $t/R$.

With $|\delta\mathbf m|^2 = f(r)^2\sin^2(\Phi_0-\beta)$, the thickness
integral gives $t$ and the azimuthal integral gives
$\int_0^{2\pi}\sin^2(\Phi_0-\beta)\,d\theta=\pi$, yielding
\begin{equation}
\delta E_\mathrm{ms}[f]
\approx
\mathcal N\pi\mu_0M_\text{s}^2tR^2
\int_0^1
f(\eta)^2\,\eta\,d\eta.
\label{eq:supp_Ems_local}
\end{equation}
For a deformation of constant amplitude,
Eq.~(\ref{eq:supp_Ems_local}) reduces to
$\tfrac12\mu_0\mathcal{N}M_\text{s}^2 f^2 V$, the energy of a
body of volume $V = \pi R^2 t$ uniformly magnetized with demagnetizing factor $\mathcal{N}$. In this sense $\mathcal{N}$ is the
effective in-plane demagnetizing factor of the $m=1$ deformation,
which motivates the uniform-cylinder estimate adopted in the main
text; its value is fixed in Sec.~\ref{sec:supp_N}.

\section{Radial profile: Euler--Lagrange equation and solution}
\label{sec:supp_radial}

\subsection{Euler--Lagrange equation}

Collecting the three deformation contributions
Eqs.~(\ref{eq:supp_dEex}), (\ref{eq:supp_dEZ}) and
(\ref{eq:supp_Ems_local}), the deformation energy is a single radial
functional $E[f] = \int_0^1 \mathcal{L}(\eta,f,f')\,d\eta$ with
\begin{equation}
\mathcal{L} = \pi t\Big[ A\big(\eta f'^2 + \tfrac{f^2}{\eta}\big)
            + \mu_0 M_\text{s} H R^2\, \eta f
            + \mathcal{N}\mu_0 M_\text{s}^2 R^2\,\eta f^2 \Big].
\label{eq:supp_lagrangian}
\end{equation}
With $\partial\mathcal{L}/\partial f' = 2\pi t A\,\eta f'$ and
$\partial\mathcal{L}/\partial f = \pi t[\,2A f/\eta
+ \mu_0 M_\text{s} H R^2\eta + 2\mathcal{N}\mu_0 M_\text{s}^2 R^2\eta f\,]$, the
Euler--Lagrange equation
$\frac{d}{d\eta}(\partial\mathcal{L}/\partial f')
= \partial\mathcal{L}/\partial f$ gives, after dividing by $\pi t$ and
multiplying by $\eta/(2A)$,
\begin{equation}
\eta^2 f'' + \eta f' - f
- \frac{\mathcal{N}\mu_0 M_\text{s}^2 R^2}{A}\,\eta^2 f
= \frac{\mu_0 M_\text{s} H R^2}{2A}\,\eta^2 .
\label{eq:supp_EL_dim}
\end{equation}
In terms of the exchange length $\ell_\text{ex}=\sqrt{2A/(\mu_0 M_\text{s}^2)}$
and the two dimensionless groups
\begin{equation}
\mathcal{A}^2 = \frac{A}{\mathcal{N}\mu_0 M_\text{s}^2 R^2}
             = \frac{\ell_\text{ex}^2}{2\mathcal{N}R^2},
\qquad
\mathcal{B} = \frac{\mu_0 M_\text{s} H R^2}{2A} = \frac{H}{H_\text{ex}},
\label{eq:supp_AB}
\end{equation}
with $H_\text{ex}=2A/(\mu_0 M_\text{s} R^2)$, this is the governing equation of
the main text, its Eq.~(12),
\begin{equation}
\eta^2 f''(\eta) + \eta f'(\eta)
- \Big(1 + \frac{\eta^2}{\mathcal{A}^2}\Big) f(\eta)
= \mathcal{B}\,\eta^2 ,
\label{eq:supp_ODE}
\end{equation}
on $\eta\in[0,1]$. Both boundary conditions follow from physics rather
than convenience.

\paragraph{Core condition $f(0)=0$.}
Near the origin the magnetostatic term $\eta^2/\mathcal{A}^2$ is
negligible against unity, so the homogeneous part of
Eq.~(\ref{eq:supp_ODE}) reduces to the Euler equation
$\eta^2 f'' + \eta f' - f = 0$; the power law $f=\eta^s$ gives
$s^2-1=0$, hence $f = C_1\,\eta + C_2\,\eta^{-1}$. The divergent branch
makes the deformation angle $\delta\Phi = f\sin(\Phi_0-\beta)$ blow up at
the vortex center and gives an exchange density $f^2/\eta\sim\eta^{-3}$
in Eq.~(\ref{eq:supp_dEex}) with divergent integral; finiteness of the
energy therefore forces $C_2=0$, leaving $f\propto\eta$ near the origin
and hence $f(0)=0$.

\paragraph{Edge condition $f'(1)=0$.}
Integrating the first variation of $E[f]$ by parts leaves, besides the
bulk Euler--Lagrange equation, the boundary term
\begin{equation}
\delta E\big|_\text{bdy} = 2\pi A t\,\big[\eta\,f'(\eta)\,\delta f(\eta)\big]_0^1 .
\end{equation}
It vanishes at $\eta=0$ by core regularity; at the rim the magnetization
angle is not pinned by any external constraint, so $\delta f(1)$ is
arbitrary and stationarity forces the natural free-edge condition
$f'(1)=0$. Physically, no radial torque acts at the free edge, so the
profile meets the rim with zero slope.

\subsection{General solution and closed-form approximation}

In the core coordinate $\xi=\eta/\mathcal{A}$, Eq.~(\ref{eq:supp_ODE}) is
the inhomogeneous modified Bessel equation of order one,
\begin{equation}
\xi^2 f_{\xi\xi} + \xi f_\xi - (1+\xi^2)\,f = \mathcal{B}\mathcal{A}^2\,\xi^2 ,
\label{eq:supp_bessel}
\end{equation}
so that $\mathcal{A}$ is absorbed into the radial variable and the whole
radial structure is set by the crossover at $\xi\sim1$, i.e.\ by the
physical length $\ell_\text{sat}=\mathcal{A}R$.

Its homogeneous solutions are the modified Bessel functions $I_1(\xi)$ and
$K_1(\xi)$~\cite{sm:abramowitz1965}, and in $f=f_\text{p}+C_1I_1+C_2K_1$ the
boundary conditions suppress both: core regularity forbids the $1/\xi$
divergence of $K_1$, so $C_2=0$, while the free edge sits at
$\xi=1/\mathcal{A}\gg1$, where $I_1$ is already exponentially large, so
$C_1=O(e^{-1/\mathcal{A}})$. The profile is carried by the particular
solution alone; that solution is not elementary, being expressible in
terms of the modified Struve function as
$f=\mathcal{B}\mathcal{A}^2\,(\pi/2)\left[L_1(\xi)-I_1(\xi)\right]$~\cite{sm:abramowitz1965},
and is evaluated numerically here.

Its physical content is two dominant balances separated by $\xi\sim1$. For
$\xi\ll1$ ($r\ll\ell_\text{sat}$) the $\xi^2$ terms drop and the Euler
equation met above gives $f\propto\xi$: exchange stiffly suppresses the
deformation near the core, which rises linearly from $f(0)=0$. For
$\xi\gg1$ ($r\gg\ell_\text{sat}$) the derivative terms drop and
$-\xi^2f=\mathcal{B}\mathcal{A}^2\xi^2$ gives the plateau
$f\to f_\text{bulk}=-\mathcal{B}\mathcal{A}^2=-H/(2\mathcal{N}M_\text{s})$.

\begin{figure}[htbp]
    \centering
    \includegraphics[width=0.95\linewidth]{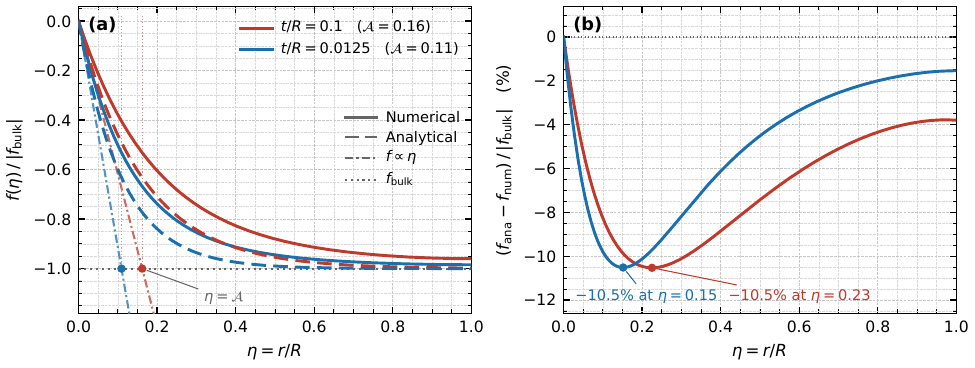}
    \caption{Radial deformation profile for the two benchmark geometries,
    $t/R = 0.1$ (red, $\mathcal{A}=0.16$) and $t/R = 0.0125$ (blue,
    $\mathcal{A}=0.11$), with $\mathcal{N}$ from Eq.~(\ref{eq:supp_N_sato}).
    Equation~(\ref{eq:supp_ODE}) is linear in $\mathcal{B}$, so both panels are
    independent of the applied field. (a)~Profile normalized to the plateau,
    $f(\eta)/|f_\text{bulk}|$. Solid lines: numerical solution of
    Eq.~(\ref{eq:supp_ODE}) (``Numerical''). Dashed lines: closed-form
    single-exponential interpolant~(\ref{eq:supp_closed}) (``Analytical'').
    The two asymptotes are also shown: the interpolant's inner linear
    branch $f\propto\eta$ (dash-dotted) and the outer plateau
$f_\text{bulk}=-H/(2\mathcal{N}M_\text{s})$ reached far from it (dotted).
    The vertical dotted lines mark the crossover at $\eta=\mathcal{A}$, i.e.\ the
    saturation length $\ell_\text{sat}=\mathcal{A}R$, where the two asymptotes
    intersect (dots). (b)~Signed deviation of the interpolant from the numerical
    solution, in percent of $|f_\text{bulk}|$. The interpolant reproduces the
    plateau and the crossover scale, but lies below the exact profile across
    most of the radius, by up to $10.5\%$ of the plateau value (at
$\eta\approx0.23$ for $t/R=0.1$ and $\eta\approx0.15$ for $t/R=0.0125$), and
    is still $3.9\%$ and $1.7\%$ off, respectively, at $\eta=0.9$.}
    \label{fig:supp_ode}
\end{figure}

The simplest elementary function carrying this content is the
single-exponential interpolant of the main text, its Eq.~(15)~\cite{sm:bender1999},
\begin{equation}
f(\eta) = -\frac{H}{2\mathcal{N}M_\text{s}}\big(1 - e^{-\eta/\mathcal{A}}\big),
\label{eq:supp_closed}
\end{equation}
which reproduces both limiting values exactly, the crossover length
$\ell_\text{sat}$, and the two properties the ansatz relies on: linearity
in $H$ and $f<0$, so that the deformation opposes the field-driven angular
excursion. It approximates the exact profile rather than equalling it: the
two coincide on the outer plateau, where the ansatz is evaluated over most
of the disk, and differ across the crossover region, which for the
benchmark geometries is not thin: the interpolant runs up to $10.5\%$ of
the plateau value too negative in both, peaking at $\eta\approx0.23$ for
$\mathcal{A}=0.16$ ($t/R=0.1$) and at $\eta\approx0.15$ for
$\mathcal{A}=0.11$ ($t/R=0.0125$), and still $3.9\%$ and $1.7\%$ off,
respectively, at $\eta=0.9$ (Fig.~\ref{fig:supp_ode}). Because
Eq.~(\ref{eq:supp_bessel}) depends on $\eta$ only through
$\xi=\eta/\mathcal{A}$, this peak deviation is a geometry-independent
property of the interpolant, reached at $\eta\simeq1.39\,\mathcal{A}$.
The interpolant's inner slope is itself approximate: near the core the
exact profile rises as $-(\pi/4)\,\mathcal{A}\mathcal{B}\,\eta$, a factor
$\pi/4$ of the interpolant's initial slope $-\mathcal{A}\mathcal{B}\,\eta$
(the interpolant is $4/\pi\approx1.27$ times steeper), so the exact inner
asymptote meets the plateau at $\eta=4\mathcal{A}/\pi$ rather than at
$\eta=\mathcal{A}$. It also meets the
free-edge condition $f'(1)=0$ derived above to exponential accuracy:
$f'(1) = -\big(H/2\mathcal{N}M_\text{s}\big)\,\mathcal{A}^{-1}e^{-1/\mathcal{A}}$,
a residual edge slope of about $1\%$ of the plateau value at
$\mathcal{A}=0.16$.

\section{Magnetostatic demagnetizing factor of the \texorpdfstring{$m=1$}{m=1} deformation}
\label{sec:supp_N}

The coefficient $\mathcal{N}$ of the local
reduction~(\ref{eq:supp_Ems_local}) is the effective in-plane
demagnetizing factor of the deformed texture and depends only on the
disk aspect ratio $t/R$. The main text adopts, as its Eq.~(11), the algebraic estimate
of Sato and Ishii~\cite{sm:sato1989} for the in-plane demagnetizing
factor of a uniformly magnetized cylinder,
\begin{equation}
\mathcal{N}_\text{Sato}(t/R) = \frac{t/R}{\sqrt{\pi} + 2\,t/R}.
\label{eq:supp_N_sato}
\end{equation}
Equation~(\ref{eq:supp_N_sato}) is derived for a body magnetized
uniformly along a single axis~\cite{sm:sato1989}, and is itself an algebraic
estimate of that uniform-cylinder factor; the $m=1$ deformation is
moreover not uniform. Its accuracy for the present problem therefore
cannot be assumed. This section fixes $\mathcal{N}$ by direct
micromagnetic calibration and quantifies the deviation from
Eq.~(\ref{eq:supp_N_sato}).

\subsection{Numerical calibration}

We calibrate $\mathcal{N}$ directly against
the magnetostatic solver of MuMax+, using only frozen analytical
textures and no magnetization relaxation. For a given geometry
$(R,t)$ and material $(M_\text{s}, A)$, two configurations are submitted to
the solver: the ideal centered vortex $\Phi_0$ and the
DIVA-perturbed profile $\Phi_0 + f(\eta;\mathcal{N})\sin(\Phi_0-\beta)$
built from the analytical radial profile $f(\eta;\mathcal{N})$ of the
main text, its Eq.~(15). Their magnetostatic energy difference,
\begin{equation}
\begin{split}
\Delta E_\text{ms}(\mathcal{N}) \equiv{}&
  E_\text{ms}\!\left[\Phi_0 + f(\eta;\mathcal{N})\sin(\Phi_0-\beta)\right]\\
  &- E_\text{ms}[\Phi_0],
\end{split}
\label{eq:supp_dEms}
\end{equation}
is the exact magnetostatic cost of the deformation. Equating it to the
right-hand side of Eq.~(\ref{eq:supp_Ems_local}) yields the calibrated
value
\begin{equation}
\mathcal{N} =
\frac{\Delta E_\text{ms}(\mathcal{N})}
{\pi\mu_0 M_\text{s}^2\, t R^2 \int_0^1 f(\eta;\mathcal{N})^2\,\eta\, d\eta}.
\label{eq:supp_N_extract}
\end{equation}
Because the trial profile $f(\eta;\mathcal{N})$ itself depends on
$\mathcal{N}$, Eq.~(\ref{eq:supp_N_extract}) is a fixed-point problem,
solved by the self-consistent iteration of
Algorithm~\ref{alg:supp_N}.

\begin{algorithm}[htbp]
\caption{Self-consistent calibration of the in-plane demagnetizing factor
$\mathcal{N}$ for a single geometry.}
\label{alg:supp_N}
\begin{algorithmic}[1]
\Require geometry $(R,t)$, material $(M_\text{s},A)$, reference field
  $H=\SI{1}{\milli\tesla}$, tolerance $\varepsilon=10^{-3}$
\Ensure calibrated demagnetizing factor $\mathcal{N}^\ast$
\State $E_0 \gets E_\text{ms}[\Phi_0]$
  \Comment{frozen ideal vortex; evaluated once}
\State $\mathcal{N} \gets \mathcal{N}_\text{Sato}(t/R)$
  \Comment{initial guess, Eq.~(\ref{eq:supp_N_sato})}
\Repeat
  \State $\mathcal{N}_\text{old} \gets \mathcal{N}$
  \State build $f(\eta;\mathcal{N}_\text{old})$ and the frozen profile
    $\Phi_0 + f(\eta;\mathcal{N}_\text{old})\sin(\Phi_0-\beta)$
  \State $E_1 \gets E_\text{ms}\!\left[\Phi_0 + f(\eta;\mathcal{N}_\text{old})\sin(\Phi_0-\beta)\right]$
    \Comment{MuMax+, no relaxation}
  \State $\Delta E_\text{ms} \gets E_1 - E_0$
    \Comment{Eq.~(\ref{eq:supp_dEms})}
  \State $\displaystyle \mathcal{N} \gets
    \frac{\Delta E_\text{ms}}{\pi\mu_0 M_\text{s}^2\, t R^2
    \int_0^1 f(\eta;\mathcal{N}_\text{old})^2\,\eta\,d\eta}$
    \Comment{Eq.~(\ref{eq:supp_N_extract})}
\Until{$|\mathcal{N}-\mathcal{N}_\text{old}|/\mathcal{N}_\text{old} < \varepsilon$}
\State \Return $\mathcal{N}^\ast \gets \mathcal{N}$
\end{algorithmic}
\end{algorithm}

\noindent The iteration runs at a fixed reference field
$H=\SI{1}{\milli\tesla}$, within the linear-response regime where
$\Delta E_\text{ms}\propto H^2$ so that the extracted $\mathcal{N}$ is
field-independent. The solver enters only as a numerical evaluator of a
single energy integral on a prescribed angle configuration; the
procedure is external to the analytical derivation and is performed
once per geometry.

\subsection{Results and comparison with the uniform estimate}

\begin{figure}[htbp]
    \centering
    \includegraphics[width=0.95\linewidth]{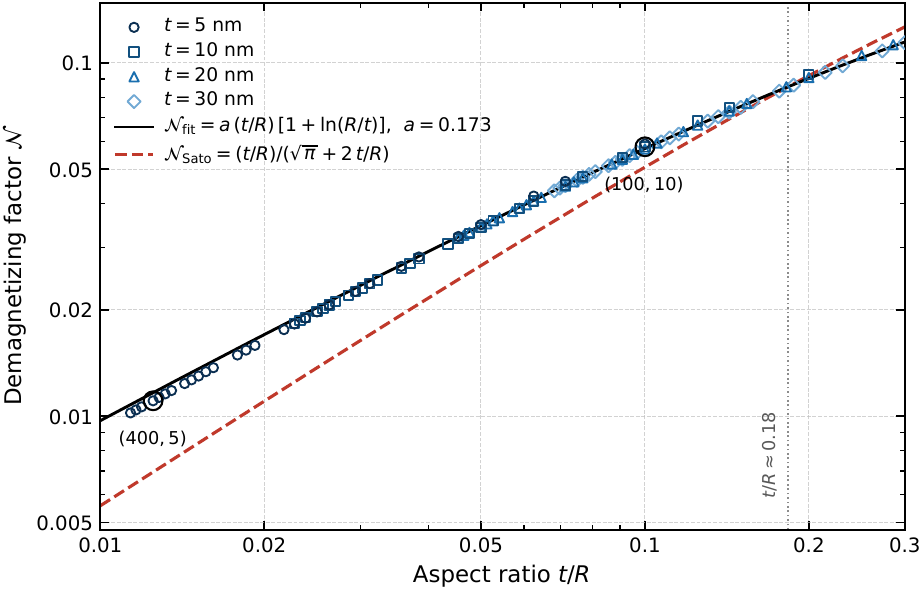}
    \caption{In-plane demagnetizing factor $\mathcal{N}$ of the $m=1$
    deformation as a function of the disk aspect ratio $t/R$. Markers:
    values calibrated for Permalloy, grouped by disk thickness
    ($t = 5, 10, 20, 30$~nm; the 116 converged runs of the 160-geometry
    sweep are shown, see Sec.~S6.2),
    from frozen-configuration MuMax+ magnetostatic energies via
    Eqs.~(\ref{eq:supp_dEms})--(\ref{eq:supp_N_extract}), obtained with the
    self-consistent iteration of Algorithm~\ref{alg:supp_N}. Solid line:
    effective thin-film vortex fit Eq.~(\ref{eq:supp_N_fit}) with
    $a = 0.173$. Dashed line: uniform-cylinder demagnetizing factor of Sato
    and Ishii~\cite{sm:sato1989}, Eq.~(\ref{eq:supp_N_sato}). The calibrated
    factor exceeds the algebraic estimate at small $t/R$ and crosses it near
    $t/R \approx 0.18$ (vertical line). The two benchmark geometries of the
    main text are circled and labelled $(R, t)$ in nanometres. The four
    thickness series superpose onto a single curve, showing that
    $\mathcal{N}$ depends on $t$ and $R$ only through $t/R$.}
    \label{fig:supp_msfactor}
\end{figure}

Figure~\ref{fig:supp_msfactor} reports the calibrated demagnetizing factor
$\mathcal{N}$ as a function of the aspect ratio $t/R$, together with the
uniform-cylinder estimate~(\ref{eq:supp_N_sato}). The numerical points
(``Py'') lie systematically above the Sato curve at small
$t/R$ and cross it near $t/R \approx 0.18$. For the two geometries
used in the main text,
\begin{equation}
\begin{aligned}
(R,t)&=(100,10)~\text{nm}: &&\mathcal{N}\approx 0.058,\ \mathcal{N}_\text{Sato}\approx 0.051,\\
(R,t)&=(400,5)~\text{nm}: &&\mathcal{N}\approx 0.011,\ \mathcal{N}_\text{Sato}\approx 0.0070,
\end{aligned}
\end{equation}
The calibrated values are converged to a relative tolerance $\varepsilon = 10^{-3}$ (Algorithm~\ref{alg:supp_N}).
i.e.\ Eq.~(\ref{eq:supp_N_sato}) underestimates the calibrated
coefficient by ${\sim}12\%$ in the $t/R = 0.1$ disk and by ${\sim}37\%$ in
the $t/R = 0.0125$ disk.

The calibrated points are well described by the one-parameter
logarithmic form
\begin{equation}
\mathcal{N}_\text{fit}(t/R) = a\,\frac{t}{R}\left[1 + \ln\!\left(\frac{R}{t}\right)\right],
\qquad a \approx 0.173.
\label{eq:supp_N_fit}
\end{equation}
The single fit parameter $a$ collapses the numerical data to within
${\sim}5\%$ across more than a decade in $t/R$ (solid line in
Fig.~\ref{fig:supp_msfactor}); the functional form
$(t/R)\left[1+\ln(R/t)\right]$ is the logarithmic scaling familiar
from thin-film magnetostatics, which Wolf et al.~\cite{sm:wolf2007}
report for thin square platelets and which we find equally describes
the disk geometry.

\section{Total-energy comparison: independent contributions and both geometries}
\label{sec:supp_energy_decomp}

The main text compares only the total micromagnetic energy of the
displaced vortex for the $t/R = 0.1$ disk, not the individual
contributions. Here we resolve the comparison into
the independent energy contributions (exchange, Zeeman and
magnetostatic) and extend it to the $t/R = 0.0125$ disk. Each energy is obtained
by evaluating the micromagnetic energy functional directly on the frozen
analytical configurations, without further relaxation.

\subsection{Aspect ratio $t/R = 0.1$}

\begin{figure}[htbp]
    \centering
    \begin{minipage}{0.49\linewidth}\centering
        \includegraphics[width=\linewidth]{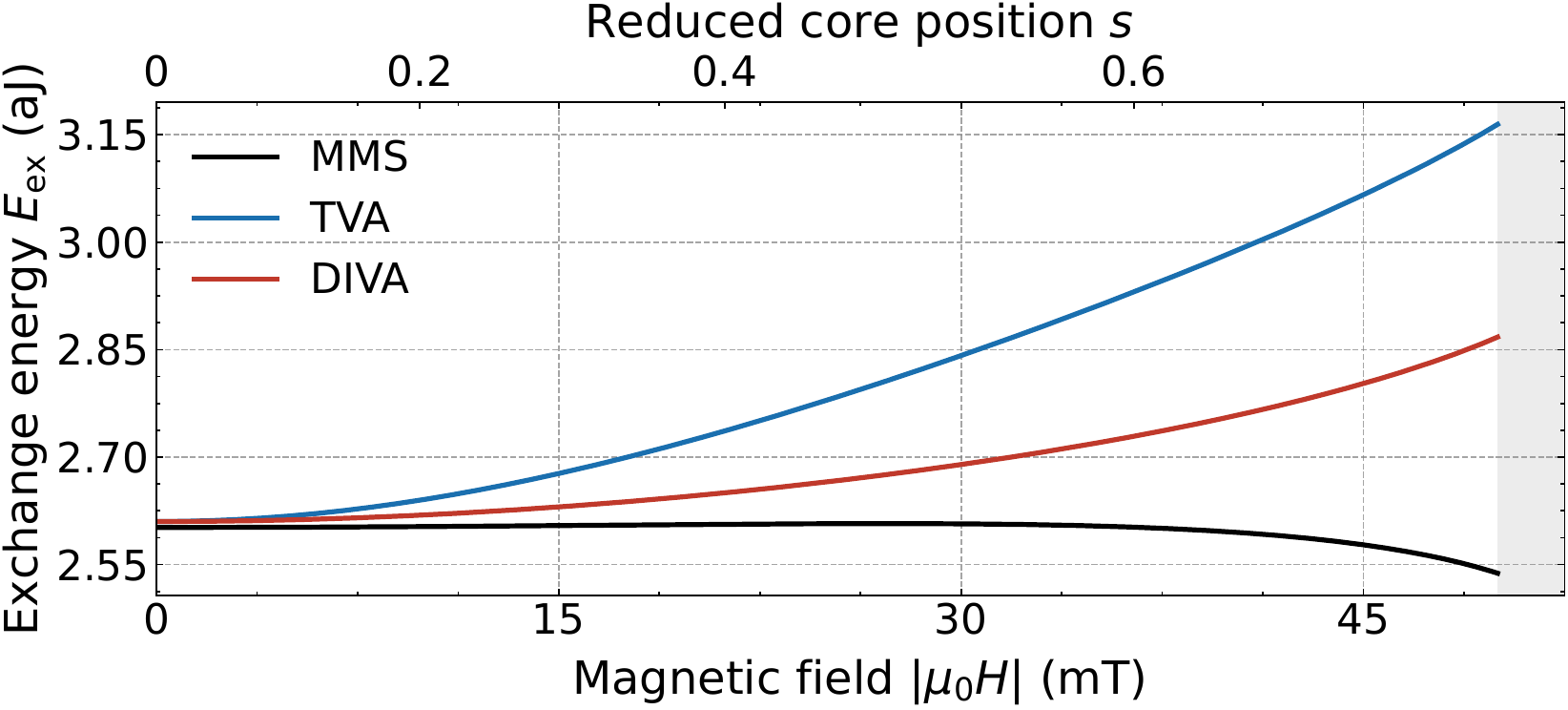}
    \end{minipage}\hfill
    \begin{minipage}{0.49\linewidth}\centering
        \includegraphics[width=\linewidth]{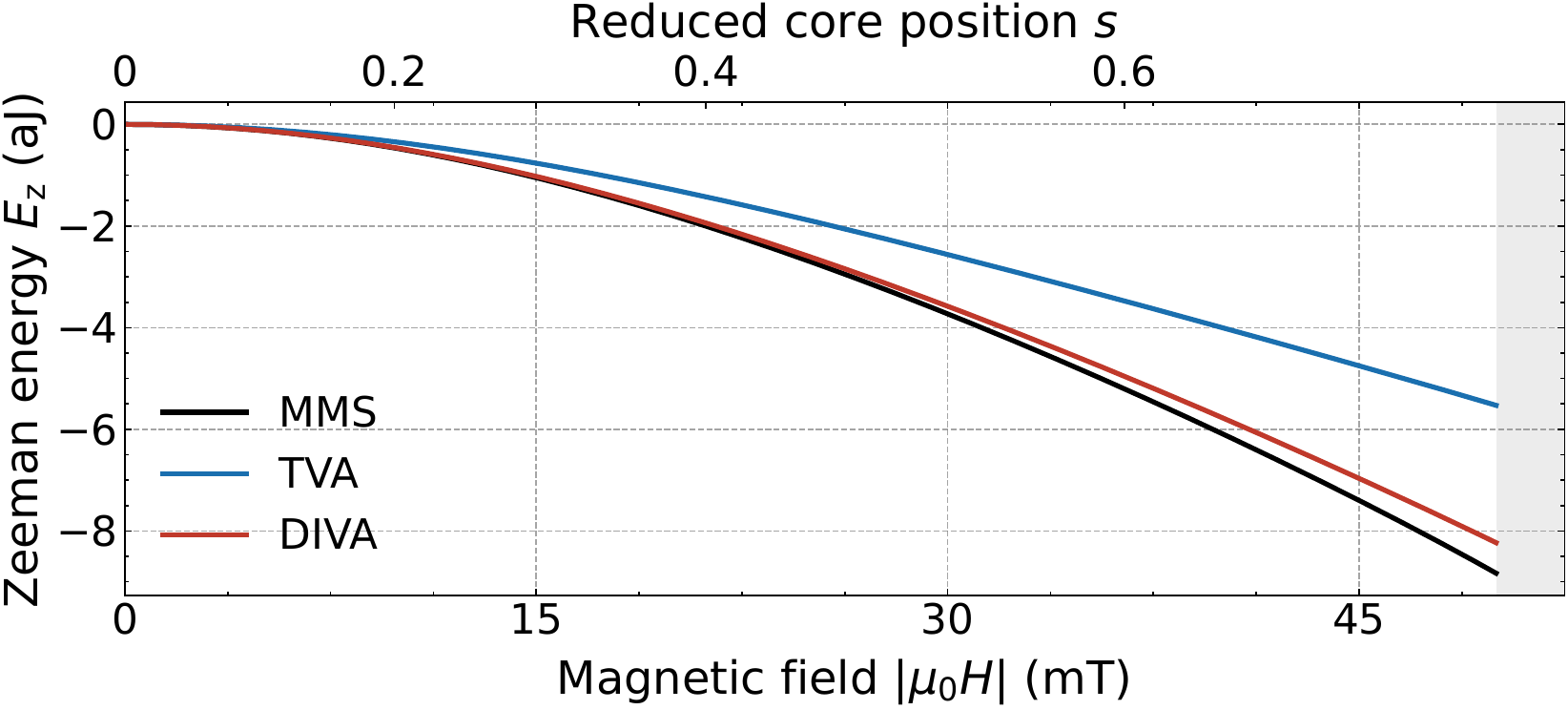}
    \end{minipage}\\[0.6em]
    \begin{minipage}{0.49\linewidth}\centering
        \includegraphics[width=\linewidth]{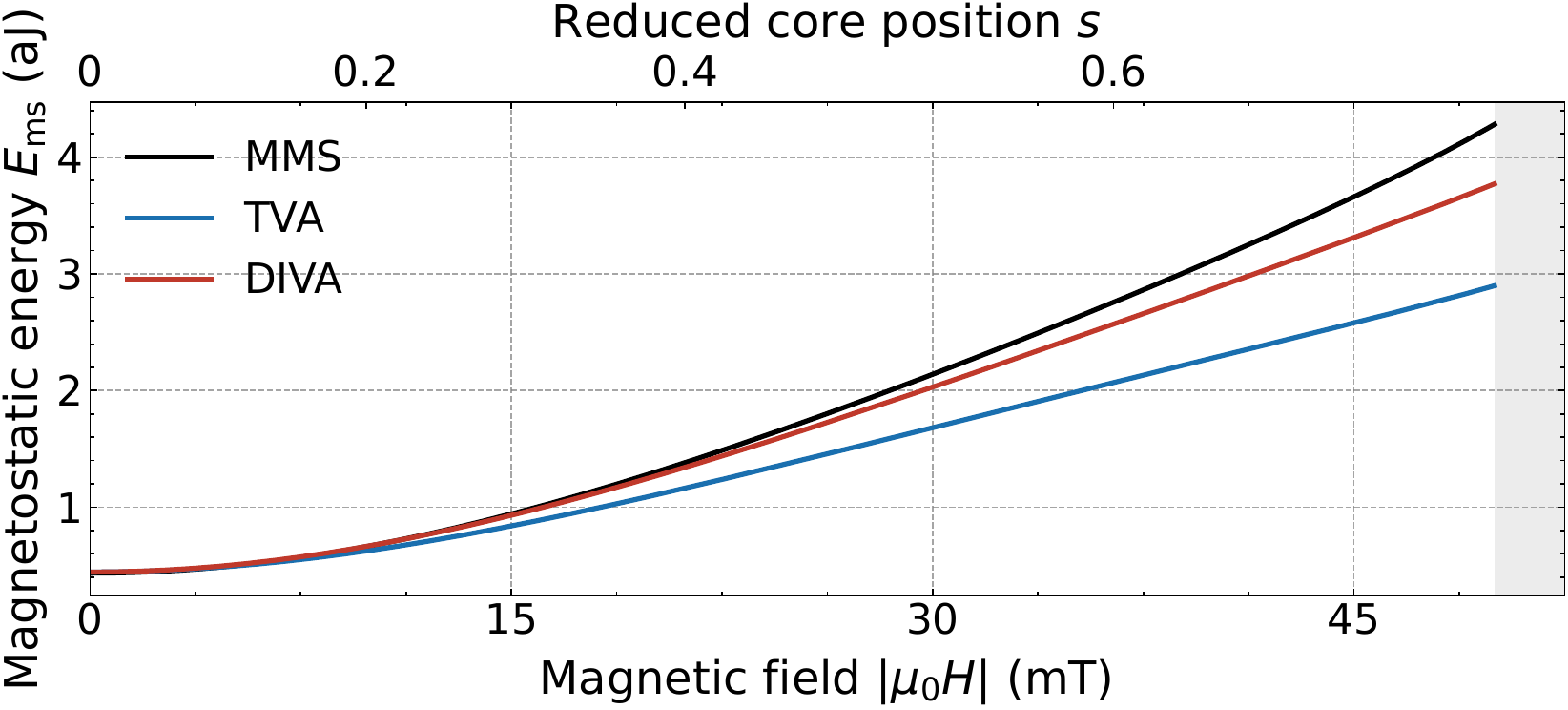}
    \end{minipage}\hfill
    \begin{minipage}{0.49\linewidth}\centering
        \includegraphics[width=\linewidth]{figs/fig4_energy_total_R100_t10.pdf}
    \end{minipage}
    \caption{Independent energy contributions of the displaced vortex for
    the $t/R = 0.1$ disk ($R=\SI{100}{\nano\metre}$, $t=\SI{10}{\nano\metre}$,
    Permalloy parameters, $\beta=0$, $C=+1$) as a function of the applied
    field $|\mu_0 H|$ (bottom axis of each panel) and of the associated
    reduced core position $s$ (top axis): exchange $E_\text{ex}$ (top left),
    Zeeman $E_Z$ (top right), magnetostatic $E_\text{ms}$ (bottom left), and
    total $E_\text{tot}$ (bottom right). Each panel compares the full
    micromagnetic simulation (MMS, black), the TVA (blue), and DIVA (red),
    the latter two evaluated at the micromagnetic core position. In each
    panel, the shaded band marks $s>1-\mathcal{A}$, beyond the validity
    window of Eq.~(17) of the main text; for this geometry the core is
    expelled at $s\approx0.8$, before the window is reached, so the bands
    contain no data. The
    bottom-right panel reproduces Fig.~4 of the main text.}
    \label{fig:supp_energy_thick}
\end{figure}

Figure~\ref{fig:supp_energy_thick} shows the exchange, Zeeman and
magnetostatic energies together with the total, as functions of the
applied field, for the relaxed micromagnetic state (MMS), the TVA, and
DIVA, all evaluated at the micromagnetic core position. Since the core
position is common to the three configurations, the curves differ only
through the field-induced deformation. The Zeeman energy decreases as the
texture tilts toward the field: DIVA reproduces the micromagnetic decrease
through its $m=1$ deformation, whereas the deformation-free TVA
underestimates its magnitude. The TVA simultaneously overestimates the
exchange energy, while both ansätze lie below the micromagnetic
magnetostatic energy. These per-contribution deviations do not cancel:
summed, they leave the TVA total energy above the micromagnetic one by an
amount that grows quadratically with field (the energy the texture
would release by relaxing into the deformation), whereas DIVA reproduces
each contribution closely enough to track the total to within the
field-independent offset of the out-of-plane core regularization.

\subsection{Aspect ratio $t/R = 0.0125$}

\begin{figure}[htbp]
    \centering
    \begin{minipage}{0.49\linewidth}\centering
        \includegraphics[width=\linewidth]{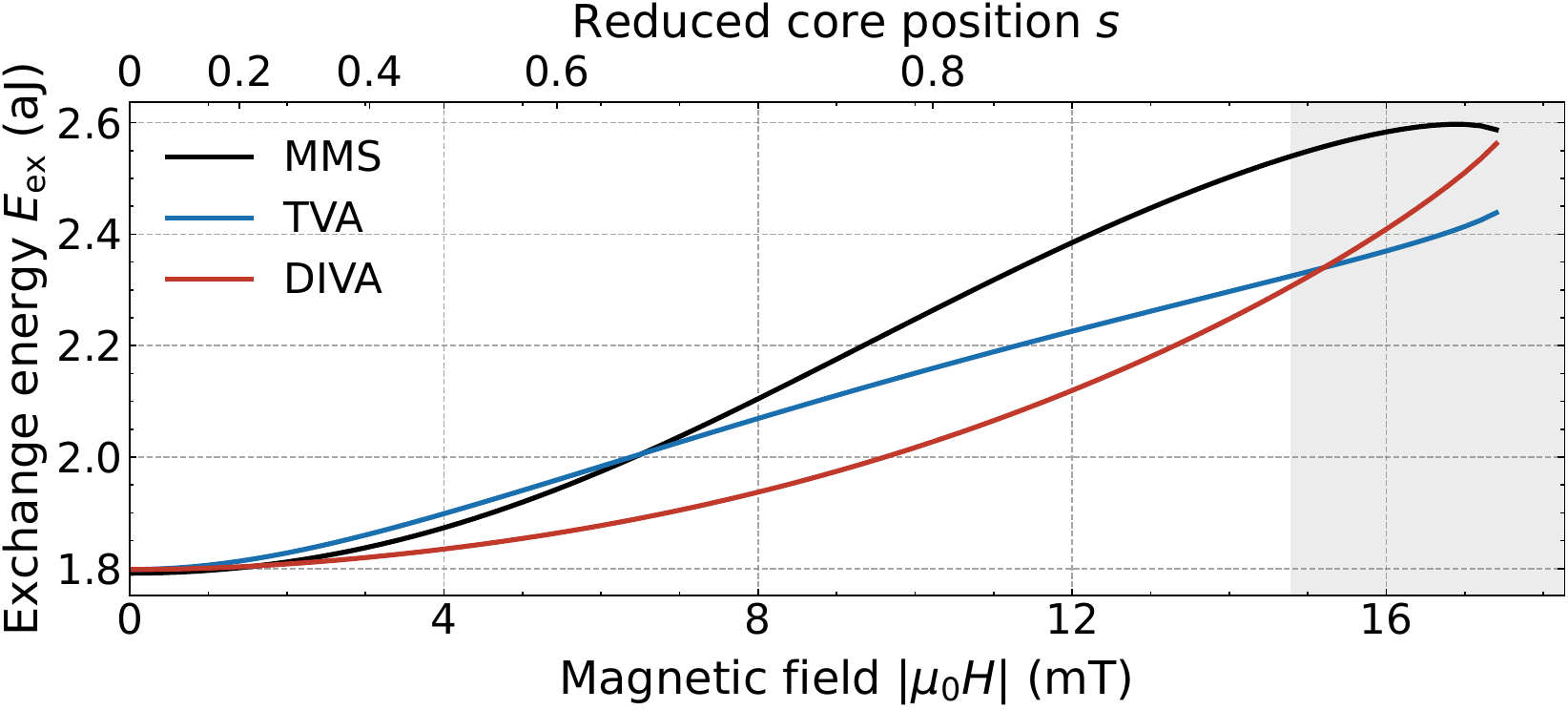}
    \end{minipage}\hfill
    \begin{minipage}{0.49\linewidth}\centering
        \includegraphics[width=\linewidth]{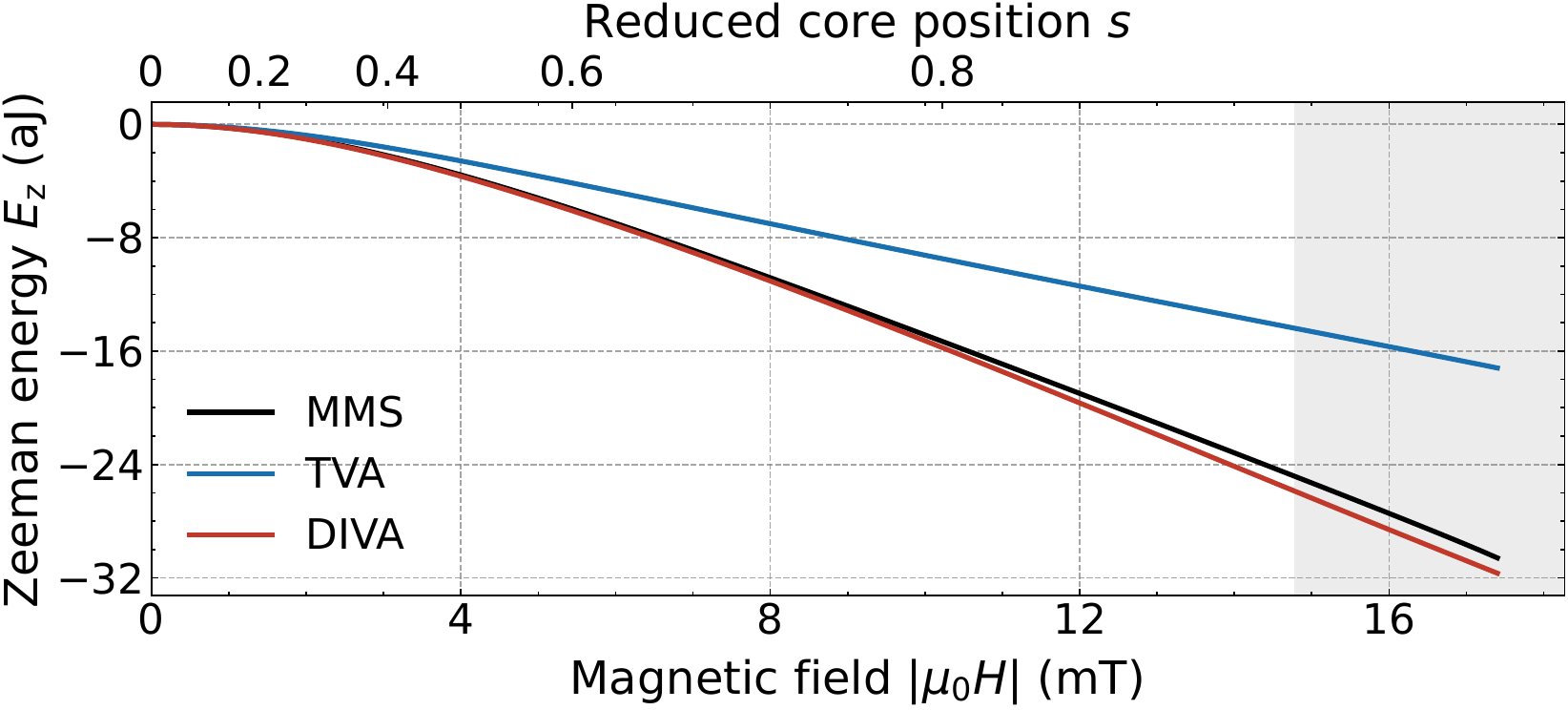}
    \end{minipage}\\[0.6em]
    \begin{minipage}{0.49\linewidth}\centering
        \includegraphics[width=\linewidth]{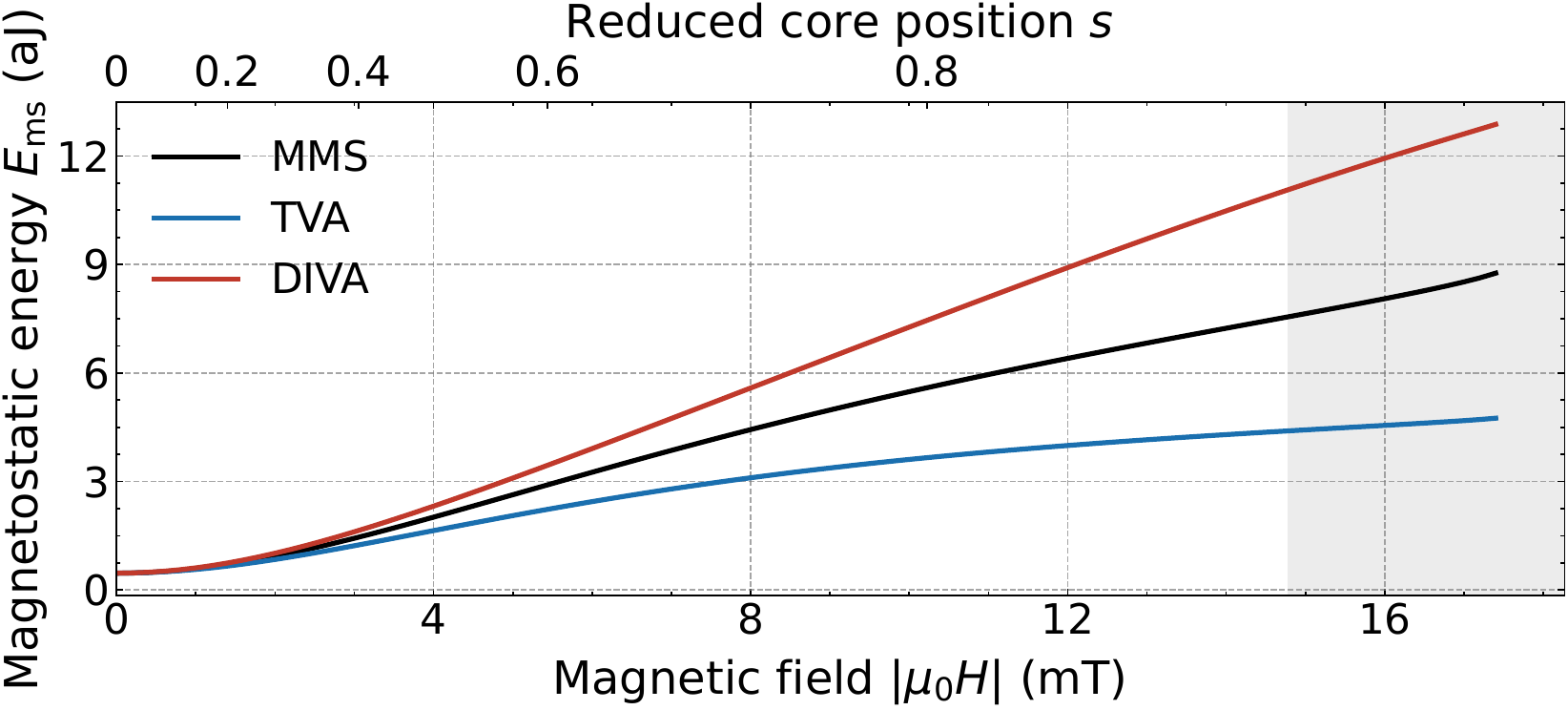}
    \end{minipage}\hfill
    \begin{minipage}{0.49\linewidth}\centering
        \includegraphics[width=\linewidth]{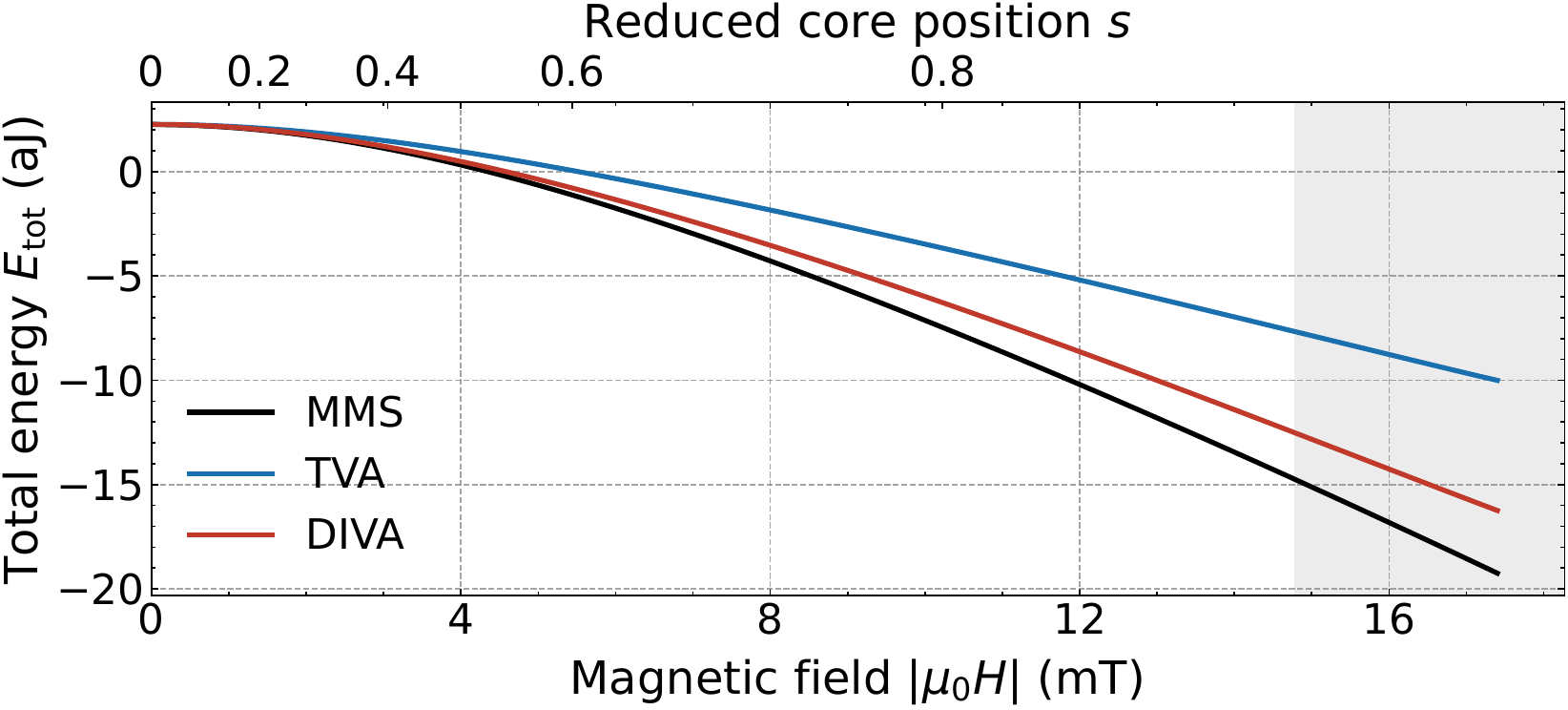}
    \end{minipage}
    \caption{Independent energy contributions of the displaced vortex for
    the $t/R = 0.0125$ disk ($R=\SI{400}{\nano\metre}$, $t=\SI{5}{\nano\metre}$,
    Permalloy parameters, $\beta=0$, $C=+1$) as a function of the applied
    field $|\mu_0 H|$ (bottom axis of each panel) and of the associated
    reduced core position $s$ (top axis): exchange $E_\text{ex}$ (top left),
    Zeeman $E_Z$ (top right), magnetostatic $E_\text{ms}$ (bottom left), and
    total $E_\text{tot}$ (bottom right). Each panel compares the full
    micromagnetic simulation (MMS, black), the TVA (blue), and DIVA (red),
    the latter two evaluated at the micromagnetic core position. In each
    panel, the shaded band marks $s>1-\mathcal{A}$, beyond the validity
    window of Eq.~(17) of the main text; unlike the $t/R=0.1$ disk, the
    core survives into this region before being expelled near the end of
    the sweep. Note the
    narrower field range than in Fig.~\ref{fig:supp_energy_thick}: the larger
    lateral extent lowers the field required to displace the core.}
    \label{fig:supp_energy_thin}
\end{figure}

The same four-panel decomposition for the $t/R = 0.0125$ disk ($R=\SI{400}{\nano\metre}$,
$t=\SI{5}{\nano\metre}$), the comparison deferred from the main text, is shown in
Fig.~\ref{fig:supp_energy_thin}. The larger lateral extent lowers the field
required to displace the core, so the comparison covers a narrower field window.

The total energy behaves as in the thicker disk: both ansätze lie above the
micromagnetic curve, TVA by ${\sim}\SI{9}{\atto\joule}$ and DIVA by
${\sim}\SI{3}{\atto\joule}$ at the end of the sweep, with the TVA deviation growing
quadratically while DIVA tracks the micromagnetic energy far more closely. The
Zeeman contribution behaves as before as well: the deformation-free TVA
underestimates the magnitude of the decrease by a wide margin, whereas DIVA
reproduces it, here slightly overshooting it.

The exchange and magnetostatic contributions, in contrast, deviate in the sense
opposite to the $t/R = 0.1$ disk, so the agreement in the total is to a
significant extent a cancellation rather than a per-contribution accuracy. DIVA
lies below the micromagnetic exchange energy over almost the whole sweep and TVA
crosses below it above ${\sim}\SI{7}{\milli\tesla}$, where in the thicker disk both
remained above; and DIVA runs above the micromagnetic magnetostatic energy
throughout, exceeding it by ${\sim}47\%$ at the largest field, where in the thicker
disk both ansätze remained below.
\newpage

\section{Micromagnetic simulation parameters}
\label{sec:supp_simparams}

All micromagnetic results reported in the main text (Figs.~2--4) and in
Secs.~S4--S5 were obtained with MuMax+~\cite{sm:mumaxplus} (Python package
\texttt{mumaxplus}, v1.1.2), run on a single NVIDIA RTX~3090 GPU on the
CISM ``Manneback'' cluster (UCLouvain). Two distinct simulation protocols
were used, described separately below: (i) fully relaxed micromagnetic
simulations (MMS) benchmarking DIVA and TVA against the core position and
energetics of the true equilibrium state
(Sec.~S6.1), and (ii) frozen-configuration
magnetostatic evaluations used only to calibrate $\mathcal{N}$
(Sec.~S6.2, cf. Sec.~\ref{sec:supp_N}).

Material parameters are those of Permalloy throughout:
$M_\text{s} = \SI{8e5}{\ampere\per\metre}$,
$A = \SI{1.07e-11}{\joule\per\metre}$, Gilbert damping $\alpha = 0.01$, and
zero magnetocrystalline anisotropy. All
simulations are athermal ($T=0$).

\subsection{Relaxed benchmark simulations (main-text Figs.~2--4, Figs.~S3--S4)}
\label{sec:supp_simparams_mms}

For each field step, the magnetization is relaxed by conjugate-gradient
energy minimization (\texttt{minimize()}). The magnetization is initialized at zero field as an analytical vortex
(chirality $C=+1$, polarity $P=+1$, core radius
$r_c=\sqrt{2}\ell_\text{ex}$) and minimized once before the field sweep
begins; the in-plane field is then ramped quasi-statically along $\beta=0$
in steps of $\Delta B = \SI{0.2}{\milli\tesla}$, each step seeded from the
relaxed magnetization of the previous step. Table~\ref{tab:supp_simparams}
lists the geometry-specific mesh parameters.

\begin{table}[htbp]
\caption{Mesh and field-step parameters for the two relaxed-MMS benchmark
geometries ($\ell_\text{ex}\approx\SI{5.16}{\nano\metre}$). The
energy-decomposition figures (main-text Fig.~4, Fig.~S3) instead use
$c_{xy}=\SI{1}{\nano\metre}$ for $R=100$~nm, $t=10$~nm.}
\label{tab:supp_simparams}
\begin{ruledtabular}
\begin{tabular}{lcc}
 & $R=100$, $t=10$~nm & $R=400$, $t=5$~nm \\
\hline
Lateral cell size $c_{xy}$ & 0.5~nm & 1~nm \\
Vertical cell size $c_z$ & 5~nm & 5~nm \\
Vertical cells $n_z$ & 2 & 1 \\
Lateral grid $n_x{=}n_y$ & 400 & 800 \\
Field step $\Delta B$ & 0.2~mT & 0.2~mT \\
\end{tabular}
\end{ruledtabular}
\end{table}

The exchange, Zeeman, magnetostatic, and total energies of Sec.~S5 are the
built-in integrated energy functions of \texttt{mumaxplus}
(\texttt{exchange\_energy}, \texttt{demag\_energy},
\texttt{zeeman\_energy} and \texttt{total\_energy}), evaluated once per
relaxed field step directly on the equilibrium state.

\subsection{Frozen-configuration \texorpdfstring{$\mathcal{N}$}{N} calibration (Sec.~S4)}
\label{sec:supp_simparams_calib}

The calibration of Sec.~\ref{sec:supp_N} uses MuMax+ purely as a
magnetostatic-energy evaluator: the analytical texture
$\Phi_0+f(\eta;\mathcal{N})\sin(\Phi_0-\beta)$ is written directly into the
magnetization (no \texttt{minimize()} or \texttt{relax()} call), and only
\texttt{demag\_energy()} is evaluated, at the fixed reference field
$H=\SI{1}{\milli\tesla}$ of Algorithm~\ref{alg:supp_N}. This protocol uses
a uniform mesh $c_x=c_y=c_z=\SI{5}{\nano\metre}$, coarser than the
relaxed-MMS meshes above, appropriate for a pure energy evaluation
repeated over the $\mathcal{N}(t/R)$ sweep of
Fig.~\ref{fig:supp_msfactor}: $R\in\{50,60,\dots,440\}$~nm (40 values)
crossed with $t\in\{5,10,20,30\}$~nm (4 values), 160 geometries in total;
the 116 runs shown as markers converged, and the remaining 44 geometries
were completed by interpolation after out-of-memory failures of the
magnetostatic solver.

\newpage
\makeatletter
\let\ARXIV@label\label
\def\label#1{\def\ARXIV@a{#1}\def\ARXIV@b{LastBibItem}%
  \ifx\ARXIV@a\ARXIV@b\else\ARXIV@label{#1}\fi}
\makeatother
%

\end{document}